\documentclass[12pt]{article}
\usepackage{amsfonts,amssymb,amsmath}
\usepackage{color}
\usepackage{theorem,cite}
\newtheorem{definition}{Definition}[section]

\newtheorem{proposition}[definition]{Proposition}
\newtheorem{property}[definition]{Property}
\newtheorem{theorem}[definition]{Theorem}

\newtheorem{lemma}[definition]{Lemma}

\theorembodyfont{\rmfamily}
\newtheorem{rmk}{Remark}[section]

\numberwithin{equation}{section}

\definecolor{brique}{rgb}{.9,.2,0}
\definecolor{blvert}{rgb}{0,.8,.85}
\definecolor{vertcl}{rgb}{0,1,.7}
\newcommand\vertcl[1]{\textcolor{vertcl}{#1}}
\newcommand\blvert[1]{\textcolor{blvert}{#1}}
\newcommand\brique[1]{\textcolor{brique}{#1}}
\def\lapth{
\begin{picture}(164,70)(0,-15)\thicklines
\put(0,0){\vertcl{\rule{20pt}{4pt}}}
\put(19,1){\vertcl{\line(1,3){23}}} 
\put(20,1){\vertcl{\line(1,3){23}}} 
\put(21,1){\vertcl{\line(1,3){23}}}
\put(22,1){\vertcl{\line(1,3){23}}}
\put(45,70){\vertcl{\line(1,-3){23}}} 
\put(44,70){\vertcl{\line(1,-3){23}}} 
\put(43,70){\vertcl{\line(1,-3){23}}}
\put(42,70){\vertcl{\line(1,-3){23}}}
\put(2,24){\vertcl{\rule{120pt}{4pt}}}
\put(65,0){\vertcl{\rule{60pt}{4pt}}}
\put(5,37){\Huge{\brique{\textbf{L}}}} 
\put(62,37){\Huge{\brique{\textbf{PTh}}}}
\put(12,-8){\blvert{\rule{92pt}{3.5pt}}}
\put(24,-15){\blvert{\rule{57pt}{3.5pt}}}
\put(36,-22){\blvert{\rule{30pt}{3.5pt}}}
\end{picture}
\raisebox{35pt}{
\begin{minipage}{320pt}\begin{center}
\textbf{Laboratoire d'Annecy-leVieux de Physique
Th\'eorique}\\[4ex]
website: \texttt{http://lappweb.in2p3.fr/lapth-2005/}
\end{center}
\end{minipage}}\\
\vspace{10pt}\quad \hrulefill\\
\vspace{10pt}}

\newcommand{\nonu}{\nonumber \\}

\newcommand{\hs}[1]{\hspace{#1 mm}}
\newcommand{\beq}{\begin{equation}}
\newcommand{\eeq}{\end{equation}}
\newcommand{\ben}{\begin{eqnarray}}
\newcommand{\een}{\end{eqnarray}}

\newcommand{\eps}{\varepsilon}

\def\cD{{\cal D}}          \def\cE{{\cal E}}          \def\cF{{\cal F}}
                    
                    \def\cL{{\cal L}}
          \def\cN{{\cal N}}          \def\cO{{\cal O}}
                    \def\cR{{\cal R}}
\def\cS{{\cal S}}                    
\def\cV{{\cal V}}          \def\cW{{\cal W}}

\def\cNb{\overline\cN}

\def\fm{{\mathfrak m}}
\def\fn{{\mathfrak n}}
\def\fp{{\mathfrak p}}
\def\fq{{\mathfrak q}}
\def\fr{{\mathfrak r}}
\def\fs{{\mathfrak s}}

\newcommand{\CC}{{\mathbb C}}
\newcommand{\EE}{{\mathbb E}}
\newcommand{\II}{{\mathbb I}}

\newcommand{\MM}{\mbox{${\mathbb M}$}}

\newcommand{\RR}{\mbox{${\mathbb R}$}}

\newcommand{\ZZ}{{\mathbb Z}}

\newcommand{\bs}[1]{{\boldsymbol{#1}}}
\newcommand{\wh}[1]{\widehat{#1}}

\newcommand{\wb}[1]{\overline{#1}}
\newcommand{\mb}[1]{\hs{4}\mbox{#1}\hs{4}}

\newcommand{\half}{\frac{1}{2}}

\newcommand{\prf}{\underline{Proof:}\ }
\newcommand{\finprf}{\null \hfill {\rule{5pt}{5pt}}\\[2.1ex]\indent}

\def\tr{\mathop{\rm tr}\nolimits}
\newcommand{\bara}{{\bar a}}

\newcommand{\bp}{{\bs p}}
\newcommand{\bq}{{\bs q}}
\newcommand{\bx}{{\bs x}}
\newcommand{\pir}{\raisebox{-.12ex}{$\stackrel{\circ}{\pi}$}{}}
\newcommand{\cWr}{\raisebox{.12ex}{$\stackrel{\circ}{\cW}$}{}}

\begin{document}

\markright{\today\dotfill DRAFT\dotfill }
\pagestyle{myheadings}
\pagestyle{empty}
\setcounter{page}{0}
\newcommand{\LAP}{LAPTH}

\hspace{-1cm}\lapth

\vspace{20mm}

\begin{center}

{\LARGE{\sffamily Universal Hubbard models
with arbitrary symmetry}}\\[1cm]

{\large G. Feverati, L. Frappat
and E. Ragoucy\footnote{
feverati@lapp.in2p3.fr, frappat@lapp.in2p3.fr, 
ragoucy@lapp.in2p3.fr}\\[.21cm] 
\textit{Laboratoire de Physique Th{\'e}orique \LAP\\[.242cm]
 LAPP, BP 110, F-74941  Annecy-le-Vieux Cedex, France. }}
\end{center}
\vfill\vfill

\begin{abstract}
We propose a general framework that leads to one-dimensional XX and Hubbard
models in full generality, based on the decomposition of an arbitrary
vector space (possibly infinite dimensional) into a direct sum of two
subspaces, the two corresponding orthogonal projectors allowing one to
define a $R$-matrix of a universal XX model, and then of a Hubbard model
using a Shastry type construction. The QISM approach ensures integrability
of the models, the properties of the obtained $R$-matrices leading to local
Hubbard-like Hamiltonians.

In all cases, the energies, the symmetry algebras and the scattering
matrices are explicitly determined. The computation of the Bethe Ansatz
equations for some subsectors of the universal Hubbard theories are
determined, while they are fully computed in the XX case.
A perturbative calculation in the large coupling regime is also done for the 
universal Hubbard models.
\end{abstract}

\vfill
\rightline{\LAP-1319/09}
\rightline{March 2009}

\newpage
\pagestyle{plain}

\section{Introduction}

The celebrated Hubbard model, introduced in the sixties
\cite{Hubbard,Gutzwiller} in order to study strongly correlated electrons,
has been widely studied since then, essentially in connection with
condensed matter physics. Due to the extent of the literature on the
subject, the reader is invited to refer to the books \cite{Monto,book} and
references therein. 
The eigenfunctions and energies of the 1D-model are known by means of the
Bethe ansatz thanks to the works of Lieb and Wu \cite{LiebWu}, the complete
set of eigenstates being obtained in \cite{EKScomp}, exploiting the $SO(4)$
symmetry present in the one-dimensional case.

The essence of the Hubbard model is rather fascinating: although the 
one-dimensional model was solved in
the late sixties, the understanding of the model in the light of the
quantum inverse scattering method became clear only twenty years after with
the works of Shastry \cite{shastry,JWshas} and Olmedilla et al.
\cite{Akutsu}. The main idea is to couple the R-matrices of two independent
XX models, through a term depending on the coupling constant of the Hubbard
potential. The complete proof of the Yang--Baxter relation for the
Hubbard R-matrix was given by Shiroishi and Wadati \cite{shiro2}.

Since then, generalizations of the Hubbard model in the framework of the
$R$-matrix formalism have been proposed. A first step was done by
Maassarani \cite{maasa2,maasa3}, extending the $R$-matrix construction to
the $gl(N)$ case.

The appearance of the Hubbard model in the context of $N=4$ super
Yang-Mills theory led to new motivations to investigate further the
supercase. The Hubbard model at half-filling, when treated perturbatively
in the coupling \cite{Rej:2005qt}, reproduces the long-ranged integrable
spin chain of Ref. \cite{Beisert:2004hm} as an effective theory. It was
conjectured in \cite{Beisert:2004hm} that the Hamiltonian of this chain be
an all-order description of the dilatation operator of $N=4$ super
Yang-Mills in the $su(2)$ subsector. There may be the possibility that some
integrable extension of the Hubbard model (e.g. involving superalgebras)
could be put in relation to other subsectors of the $N=4$ super Yang-Mills
theory.

The Hubbard model has also arisen in \cite{beisert:2006}, where an
$S$-matrix for a long-range interacting integrable quantum spin chain with
centrally extended $su(2|2)$ symmetry was constructed. This $S$-matrix was
shown to be proportional to Shastry's $R$-matrix up to a dressing phase.
This phase indeed leads to a breakdown of the conjecture of
\cite{Beisert:2004hm} beyond three loops and to transcendantal
contributions to the dilatation operator eigenvalues. However, the proposal
of a string Bethe ansatz and the appearance of the Hubbard $R$-matrix in
the study of integrable structures in view of the AdS/CFT correspondence 
ask for learning more about generalized Hubbard models. 
Hopefully, the statistical mechanics community may also find interest in
exploring these structures.

A superalgebraic generalization of the Hubbard model in the spirit of
Shastry's construction has been proposed by the authors in \cite{XX} and 
\cite{XXconf}, where
a general approach to constructing a number of super Hubbard models was
developed. Each of the obtained models can be treated perturbatively and
thus gives rise to an integrable long-ranged spin chain as an effective
theory. The symmetry of the super Hubbard model based on $gl(\fm|\fn)$ was
shown to be $gl(\fm-1|\fn-1) \oplus gl(1|1) \oplus gl(\fm-1|\fn-1) \oplus
gl(1|1)$. In this paper, we propose a general framework that leads to XX
and Hubbard models in full generality. It may also constitute an
interesting starting point for dealing with integrable bosonic Hubbard
models. More precisely, it is based on the decomposition of an arbitrary
vector space (possibly infinite dimensional) into a direct sum of two
subspaces, the two corresponding orthogonal projectors allowing one to
define a $R$-matrix of a universal XX model, and then of a Hubbard model
using a Shastry type construction. The QISM approach ensures the
integrability of the models, the properties of the obtained $R$-matrices
leading to local Hubbard-like Hamiltonians. In the finite dimensional case,
they can be interpreted in terms of `electrons' after a Jordan--Wigner
transformation \cite{JW} (see some examples in \cite{XX}).

The plan of the paper is as follows. In section \ref{sect:univXX}, we
extend the construction of XX models for algebras \cite{maasa} and
superalgebras \cite{XX} to the case of an arbitrary vector space, possibly
infinite dimensional. We focus to the general case $gl(\fm|\fn)$ in section
\ref{sect:glMN-XX}, in which the Hamiltonians are explicitly constructed
and the Bethe ansatz equations computed. In section \ref{sect:univHub}, we
tackle with the case of universal Hubbard models, performing in the same
way the calculation of the $R$-matrices, the transfer matrices and
corresponding Hamiltonians. For both models (XX and Hubbard), the energies
and the symmetry algebra, which is related to the choice of the projectors,
are determined, and the corresponding charges computed. Section
\ref{sect:BAEunivHub} is devoted to the Bethe ansatz equations for
universal $gl(\fm|\fn)$ Hubbard models. The computation of the scattering
matrix of the universal Hubbard model is performed and the BAE for some
subsectors of the theory are determined. In section
\ref{KSexpansion}, a perturbative treatment \textit{\`a la} Klein and Seitz
\cite{klstz} of the obtained Hubbard-like Hamiltonians is performed; second 
order and fourth order terms are presented. The last section is 
devoted to a short conclusion.
Finally, we give in Appendix \ref{sec:XXbos} some hints for progressing
towards integrable bosonic Hubbard models, and expose in Appendix
\ref{sec:twist} a twisted version of XX and Hubbard models, leading to
Hamiltonians that depend on phases that can be identified with a
Aharonov-Bohm phase.

\section{Universal XX models\label{sect:univXX}}
We generalize the construction given in \cite{maasa,martins,XX} to the
case of an arbitrary vector space $\cV$, possibly infinite
dimensional. 
We will use the standard auxiliary space notation, i.e. to any
operator $A\in
End(\cV)$, we associate the operator $A_{1}=A\otimes \II$ and
$A_{2}=\II\otimes A$ in $End(\cV)\otimes End(\cV)$. More
generally, considering equalities in $End(\cV)^{\otimes k}$,
$A_{j}$, $j=1,\ldots,k$, will act trivially in all spaces $End(\cV)$, but
the $j^{th}$ one.

To deal with superalgebras, we will also need a $\ZZ_{2}$ grading $[.]$ on
 $\cV$, such that $[v]=0$ will be associated to bosonic states,
$v\in\cV_{0}$, and $[v]=1$ to
fermionic states, $v\in\cV_{1}$. 

We will also assume the existence of a (super-)trace operator, defined on a subset of 
$End(\cV)$ and obeying cyclicity. When $\cV$ is finite dimensional, 
$dim\cV=\fn $, $End(\cV)$ is just the algebra $gl(\fn )$, so that 
the trace operator is the usual trace of $\fn \times \fn $ matrices. If 
$\cV$ is graded and finite dimensional, one deals with the supertrace.
When $\cV$ is infinite dimensional, the definition of a trace 
operator is more delicate, and one needs to verify that it exists and 
is cyclic for the operators we use. We address this problem in 
appendix \ref{sec:XXbos}.

\subsection{R-matrix\label{sec:univR-XX}}

To define the R-matrix of universal XX model, we need 
some preliminary notions. We define projectors
\begin{eqnarray}
\pi:\ \cV\to\ \cW\quad,\quad 
\wb\pi=\II-\pi:\  \cV\to\ \wb\cW \mb{with} \cV=\cW\oplus\wb\cW
\label{def:univpi}
\end{eqnarray}
and graded permutation operator 
\begin{eqnarray}
P_{12}:
\begin{cases} \cV\otimes\cV \ \to\ \cV\otimes\cV\\
u\otimes v\ \to\ (-1)^{[u][v]}\, v\otimes u
\end{cases}
\end{eqnarray}
Note that in auxiliary space notation, the action of the (graded)
permutation operator reads
\beq
P_{12}\,u_{1}\,v_{2}= u_{2}\,v_{1}\,.
\eeq
{From} these operators, one can construct an R-matrix
\begin{equation}
R_{12}(\lambda) = \Sigma_{12}\,P_{12} + \Sigma_{12}\,\sin\lambda +
(\II\otimes\II-\Sigma_{12})\,P_{12}\,\cos\lambda
\label{def:univRXX}
\end{equation}
where $\Sigma_{12}$ is built on the projection operators:
\begin{eqnarray}
\Sigma_{12} &=& 
\pi_{1}\,\wb\pi_{2}+\wb\pi_{1}\,\pi_{2} 
\label{def:univSigma}
\end{eqnarray}
It is easy to show that $\Sigma_{12}$ is also a projector in
$\cV\otimes\cV$:
$\left(\Sigma_{12}\right)^2=\Sigma_{12}$.

Let us introduce the parity operator $C$:
\begin{equation}
C = \pi-\wb\pi\,.
\label{eq:opC}
\end{equation}
It obeys $C^{2}=\II$ and is related to the R-matrix through the 
equalities
\begin{equation}
\Sigma_{12}=\half(1-C_{1}C_{2}) \mb{and}
\II\otimes\II-\Sigma_{12}=\half(1+C_{1}C_{2})
\label{eq:univSig-C}
\end{equation}
that allow us to rewrite the $R$-matrix as
\beq
R(\lambda) = \cos(\frac{\lambda}{2})\,\Big(
\cos(\frac{\lambda}{2})\,P_{12} + 
\sin(\frac{\lambda}{2})\,\II\otimes\II\Big)
-\sin(\frac{\lambda}{2})\,C_{1}\,C_{2}\Big(
\sin(\frac{\lambda}{2})\,P_{12} + 
\cos(\frac{\lambda}{2})\,\II\otimes\II\Big)\,.
\eeq

One has
\begin{theorem}\label{theo:univR-XX}
For all spaces $\cV$ and projectors $\pi$, the R-matrix (\ref{def:univRXX}) satisfies the following properties:
\begin{itemize}
\item[--]
Parity invariance:
\begin{equation}
C_1 \, C_2 \, R_{12}(\lambda) 
= R_{12}(\lambda) \,  C_1 \, C_2
\label{eq:univinvar}
\end{equation}
\item[--]
Sign transformation:
\begin{equation}
R_{12}(-\lambda) = C_1 \, R_{12}(\lambda) \, C_2
\label{eq:univantisym}
\end{equation}
\item[--]
Symmetry: 
\begin{equation}
R_{12}(\lambda) = R_{21}(\lambda) 
\label{eq:univsymm}
\end{equation}
\item[--]
Unitarity: 
\begin{equation}
R_{12}(\lambda) \, R_{21}(-\lambda) = (\cos^2 \lambda) \, \II \otimes
\II
\label{eq:univunitarity}
\end{equation}
\item[--] Regularity : 
\begin{equation} R_{12}(0)=P_{12}\end{equation}
\item[--]
Exchange relation: 
\begin{equation}
R_{12}(\lambda) \, R_{21}(\mu) = R_{12}(\mu) \, R_{21}(\lambda)
\label{eq:univexchange}
\end{equation}
\item[--]
Yang--Baxter equation (YBE): 
\begin{eqnarray}
&&R_{12}(\lambda_{12})\,R_{13}(\lambda_{13})\,R_{23}(\lambda_{23}) = 
R_{23}(\lambda_{23})\,R_{13}(\lambda_{13})\,R_{12}(\lambda_{12})
\qquad\nonu
&&\mb{where} \lambda_{ij} = \lambda_i-\lambda_j.
\label{eq:univYBE}
\end{eqnarray}
\item[--]
Decorated Yang--Baxter equation (dYBE): 
\begin{eqnarray}
&&R_{12}(\lambda'_{12})\,C_1\,R_{13}(\lambda_{13})
\,R_{23}(\lambda'_{23}) = 
R_{23}(\lambda'_{23})\,R_{13}(\lambda_{13})\,C_1
\,R_{12}(\lambda'_{12})\qquad\nonu
&&\mb{with} \lambda'_{ij} = \lambda_i+\lambda_j.
\label{eq:univdecYBE}
\end{eqnarray}
 \end{itemize}
\end{theorem}
\prf
The proof is strictly similar to the one done in \cite{XX}, the only 
needed relations being
\begin{equation}
C^2=\II\quad;\quad C_1 \, \Sigma_{12} = \Sigma_{12}\, C_1
= -\Sigma_{12}\, C_2 =-C_2\, \Sigma_{12} 
\label{eq:univpptC}
\end{equation}
and the relation (\ref{eq:univSig-C}). Let us also remark that this 
latter relation is equivalent to the relations
\begin{equation}
2\,\Sigma_{ij}\,\Sigma_{kj} = \Sigma_{ij}+\Sigma_{kj}-\Sigma_{ik}\,,\ \forall\ i,j,k
\mb{with} \Sigma_{ii}=0
\label{eq:sig}
\end{equation}
without any reference to the projectors $\pi$ and $\wb\pi$. However, a 
detailed analysis of the relations (\ref{eq:sig}) shows that $\Sigma_{12}$ 
must be of the form (\ref{eq:univSig-C}), up to conjugation.
\finprf
\begin{rmk}
When $\pi=0$ or $\pi=\II$, we get
$R_{12}=\cos(\lambda)\,P_{12}$, which also obeys all the
statements of theorem \ref{theo:univR-XX}, but leads to trivial models.
\end{rmk}

\begin{lemma}
If we denote by $R^{(\pi)}_{12}(\lambda)$ the R-matrix built on $\pi$, we have
$$
R^{(\pi)}_{12}(\lambda) =R^{(\II-\pi)}_{12}(\lambda)\,.
$$
\end{lemma}
\prf
We have $C^{(\pi)}=-C^{(\II-\pi)}$, leading to the property 
$\Sigma^{(\pi)}_{12}=\Sigma^{(\II-\pi)}_{12}$\,.\finprf

\subsection{Monodromy and transfer matrices\label{sec:XXmono}}
{F}rom the R-matrix, one constructs the ($L$ sites) monodromy matrix
\begin{equation}
\cL_{0<1\ldots L>}(\lambda) = R_{01}(\lambda)\,R_{02}(\lambda)\cdots
R_{0L}(\lambda)
\end{equation}
which obeys the relation
\begin{equation}
R_{00'}(\lambda-\mu)\, \cL_{0<1\ldots L>}(\lambda)\, 
\cL_{0'<1\ldots L>}(\mu) =
\cL_{0'<1\ldots L>}(\mu) \, \cL_{0<1\ldots L>}(\lambda)\,
R_{00'}(\lambda-\mu)\,.
\label{univRLL-XX}
\end{equation}
This relation allows us to construct an ($L$ sites) integrable XX spin
chain through the transfer matrix
\begin{equation}
t_{1\ldots L}(\lambda) = \tr_{0} \cL_{0<1\ldots L>}(\lambda) 
= \tr_{0} \Big(R_{01}(\lambda)\,R_{02}(\lambda)\cdots 
R_{0L}(\lambda)\Big)\,.
\end{equation}
Indeed, when the trace operator is well-defined on the monodromy 
matrix\footnote{For finite dimensional vector spaces, the trace 
operator is obviously always defined. For infinite dimensional 
spaces, one needs to be more careful: we will come back on this point 
in appendix \ref{sec:XXbos}.}, the relation (\ref{univRLL-XX}) implies that the transfer matrices for
different values of the spectral parameter commute
\begin{equation}
[t_{1\ldots L}(\lambda)\,,\,t_{1\ldots L}(\mu)]=0\,.
\end{equation}

Then, the XX-Hamiltonian is defined by
\begin{equation}
H=t_{1\ldots L}(0)^{-1}\, \frac{dt_{1\ldots 
L}}{d\lambda}(0)\,.
\end{equation}
Since the R-matrix is regular, $H$ is local:
\begin{equation}
H=\sum_{j=1}^{L} H_{j,j+1}\mb{with} H_{j,j+1}=
P_{j,j+1}\,\Sigma_{j,j+1}
\label{eq:univXXHam}
\end{equation}
where we have used periodic boundary conditions, i.e. identified the site
$L+1$ with the first site. 
 
\subsection{Symmetry of universal XX models}
\begin{proposition}
Let us consider a universal XX model based on a vector space $\cV$, 
with projectors $\pi\,:\, \cV\to\cW$ and $\wb\pi\,:\, \cV\to\wb\cW$.
For $\MM\in End(\cW)\oplus End(\wb\cW)$, one has
\begin{equation}
(\MM_{1}+\MM_{2})\,R_{12}(\lambda) = 
R_{12}(\lambda)\,(\MM_{1}+\MM_{2})\,.
\label{eq:symunivRxx}
\end{equation}
As a consequence, the transfer matrix also has a symmetry (super)algebra
$End(\cW)\oplus End(\wb\cW)$, with generators given by
\begin{equation}
\MM_{<1\ldots L>}=\MM_{1}+\MM_{2}+\ldots+\MM_{L},
\end{equation}
where $\MM\in End(\cW)\oplus End(\wb\cW)$.
The same is true for any Hamiltonian $H$ built on the transfer matrix.
\end{proposition}
\prf
Starting from a general  morphism $\MM\in End(\cV)$, a direct calculation shows
that when $\MM(\cW)\subset\cW$ and
$\MM(\wb\cW)\subset\wb\cW$, we have $\MM\,\pi=\pi\,\MM$ and
$\MM\,\wb\pi=\wb\pi\,\MM$ so that
 (\ref{eq:symunivRxx}) holds. The above conditions are equivalent to 
 $\MM\in End(\cW)\oplus End(\wb\cW)$.\\
As far as the transfer matrix is concerned, the proof is the 
well-known, once (\ref{eq:symunivRxx}) holds.
\finprf 

Since the choice of the projector $\pi$ fixes $\cW$ 
and $\wb\cW$, the above procedure allows us to associate to any 
symmetry (super)algebra $\cS$ a universal XX model possessing $\cS$ 
as symmetry.

The eigenstates of the transfer matrix will be also eigenstates of the 
Cartan generators of the symmetry algebra. These generators are 
given by $\MM_{aa}$, $a=1,\ldots,dim\cV=d$ (with possibly $d=\infty$). 
The corresponding charges will be noted 
$\Lambda=(\lambda_{1},\ldots,\lambda_{d})$. 
The charges $(\lambda_{1},\ldots,\lambda_{r})$, $r=\mbox{rank}\pi$, 
correspond to End$(\cW)$, while 
$(\lambda_{r+1},\ldots,\lambda_{d})$
are associated to End$(\wb\cW)$.
In the following, we will also need the fundamental weights 
\begin{equation}
\Lambda_{a}=(\underbrace{0,\ldots,0}_{a-1},1,0,\ldots,0)^t\,, \quad
a=1,\ldots,dim\cV=d\,. \label{eq:XXfundw}
\end{equation}
\section{Universal XX models based on $gl(\fm |\fn )$\label{sect:glMN-XX}}
\subsection{Hamiltonian and transfer matrix}
\paragraph{Projectors and $R$-matrix:}
We apply the above construction to the case where $\cV$ is the graded 
tensor product $\cV=\CC^{\fm |\fn }$, with
possibly $\fn =0$ to encompass the case $\cV=\CC^{\fm }$.
In the following, we note $\fs =\fn +\fm $.

The $\ZZ_{2}$ grading $[.]$ is defined on
indices $j$, such that $[j]=0$, $1\leq j\leq \fm $, will be associated to bosons 
and $[j]=1$, $\fm +1\leq j\leq \fm +\fn $ to
fermions. Accordingly, the elementary matrices $E_{ij}$ (with 1 at position
$(i,j)$ and 0 elsewhere) will have grade $[E_{ij}]=[i]+[j]$.

To define the projectors $\pi$ and $\wb\pi$, we introduce a subset 
$$
\cN\subset\ZZ_{\fs}= [1,\fs ]\cap\ZZ_{+}\,,
$$
and denote by $\cNb$ its complementary set, 
i.e. 
$$
\cN\cap\cNb=\emptyset\mb{and}\cN\cup\cNb=\ZZ_{\fs }\,. 
$$
We will also need the bosonic and fermionic `components' of $\cN$,
$$
\cN_{0}=\cN\cap\ZZ_{\fm }\,\quad \cN_{1}=\cN\setminus\cN_{0}
\mb{with} \cN_{0}\cup\cN_{1}=\cN\,.
$$
They are such that $[j]=0$ when $j\in\cN_{0}$ while $[j]=1$ when 
$j\in\cN_{1}$. 

To each set $\cN$, one associates projectors
\begin{eqnarray}
\pi^{(\cN)}=\sum_{j\in\cN} E_{jj}\quad,\quad 
\wb\pi=\II_{\fs }-\pi = \pi^{(\cNb)}
\label{def:pi}
\end{eqnarray}
Although these projectors depend on the set $\cN$, we will drop the
superscript $(\cN)$, keeping it only when several sets $\cN$ are
considered.

{From} these projectors, one constructs the R-matrix according to the 
general formulas (\ref{def:univRXX}) and (\ref{def:univSigma}).
This $R$-matrix obeys theorem \ref{theo:univR-XX}, with
 the parity matrix $C$:
\begin{equation}
C = \sum_{j\in\cN} E_{jj} - \sum_{k\in\cNb} E_{kk} 
=\pi-\wb\pi\,.
\label{eq:matC}
\end{equation}

\paragraph{Monodromy matrix and Hamiltonian:}
{From} the R-matrix, one constructs the ($L$ sites) monodromy and 
transfer matrices
following the general procedure explained in section \ref{sec:XXmono}.

Then, the XX-Hamiltonian is defined by eq. (\ref{eq:univXXHam}) with 
 two sites Hamiltonian
\begin{equation}
H_{j,j+1} = \sum_{i\in\cN} \sum_{\bar a\in\cNb} 
\Big( (-1)^{[\bar a]}\,E_{i\bar a} \otimes E_{\bar ai} 
+ (-1)^{[i]}\,E_{\bar ai} \otimes E_{i\bar a}
\Big)\,.
\end{equation}
Anticipating the Bethe ansatz analysis, one can see that
this Hamiltonian describes, apart from 
the `vacuum', 
 $\fm +\fn -1$ species of particles gathered into two subsets, so-called the 
`barred' $\bar a$, $\bar b$, \ldots and `unbarred' particles $a$, $b$, \ldots
corresponding to projectors $\wb\pi$ and $\pi=\II_{\fs }-\wb\pi$ 
respectively. The `barred' particles move as hard-core particles while 
the `unbarred' particles are displaced by the barred ones. This 
latter property is valid for a vacuum of `unbarred' type: obviously, 
one has to reverse `barred' and `unbarred' particles if the vacuum is 
chosen of `barred' type. 

\paragraph{Symmetry and number of  models:}
Obviously, without any loss of generality, one can choose 
\ben
\cN &=& \{1,2,\ldots,\fr_{0}\,;\,\fm +1,\fm +2,\ldots,\fm +\fr_{1}\}
\mb{with} \fr_{0}=|\cN_{0}|\,,\ \fr_{1}=|\cN_{1}|  \\
\cNb &=& \{\fr_{0}+1,\fr_{0}+2,\ldots,\fm \,;\,
\fm +\fr_{1}+1,\fm +\fr_{1}+2,\ldots,\fm +\fn=\fs \}
\een

{From} the property $R^{(\cN)}_{12}(\lambda) =R^{(\cNb)}_{12}(\lambda)$
and the isomorphism 
$gl(\fn |\fm )\simeq gl(\fm |\fn )$, one can impose 
the inequalities
$$
\fr_{0}=|\cN_{1}|\leq\frac{\fm +1}{2} \mb{and} \fm \geq \fn \,,
$$
leading to 
$\left(\left[\frac{\max+1}{2}\right]+1\right)\,(\min+1)$ 
different models, where we used the notation $\min=\min(\fn ,\fm )$ and $\max=\max(\fn ,\fm )$. 
 
The R-matrix admits a $gl(\fm -\fr_{0}|\fn -\fr_{1})\oplus 
gl(\fr_{0}|\fr_{1})$ symmetry
superalgebra whose generators have the form
\begin{equation}
\begin{array}{l}
\displaystyle E_{jk}\ ,\ j,k\in \cN \mb{for} gl(\fr_{0}|\fr_{1})\\
\displaystyle E_{jk}\ ,\ j,k\in \cNb\mb{for} gl(\fm -\fr_{0}|\fn -\fr_{1}).
\end{array}\label{eq:gen-glN-1}
\end{equation}
 
As a consequence, the transfer matrix also admits  $gl(\fm -\fr_{0}|\fn -\fr_{1})\oplus 
gl(\fr_{0}|\fr_{1})$
symmetry superalgebra, with generators given by 
\begin{equation}
\MM_{<1\ldots L>}=\MM_{1}+\MM_{2}+\ldots+\MM_{L},
\end{equation}
where $\MM$ is one of the generators given in (\ref{eq:gen-glN-1}).
The same is true for any Hamiltonian $H$ built on the transfer matrix.

Since the choice of the projector $\pi$ fixes the values of $\fr_{0}$ 
and $\fr_{1}$, the above procedure allows us to associate to any 
symmetry (super)algebra $\cS=gl(\fq|\fq')\oplus gl(\fm -\fq|\fn -\fq')$ 
a generalized XX model possessing $\cS$ 
as symmetry, provided the vector space $\cV=\CC^{\fm |\fn }$ we start from 
is large enough (i.e. $\fm \geq \fq$ and $\fn \geq \fq'$ to get $\cS$).
Conversely, from the vector space $\cV=\CC^{\fm |\fn }$, one can construct 
models possessing the symmetry:
\beq
gl(\fq|\fq')\oplus gl(\fm -\fq|\fn -\fq')\,,\ \fq\leq \fm  \mb{and} \fq'\leq \fn 
\eeq 

\subsection{BAEs for universal XX models\label{sec:BAE-XX}}
To get the BAEs of a model, one starts with a reference state, called 
the pseudo-vacuum, which is an eigenvector of the transfer matrix. 
The other states are constructed as `pseudo-excitations' on this 
pseudo-vacuum. 

\subsubsection{The pseudo-vacua sector}
The full space of states for the XX models is $(\cV)^{\otimes L}$: we 
 consider here the subspace $\cW_{vac}=(\cW)^{\otimes L}$. In this 
 subspace, the transfer matrix takes a simple form:
\begin{equation}
t_{XX}(\lambda)\Big|_{\cW_{vac}} = (\cos\lambda)^L\,P_{1L}P_{2L}
\ldots P_{L-1,L} + (\sin\lambda)^L\,\wb\fr 
\mb{with} \wb\fr =\mbox{rank}\wb\pi
\end{equation}
One recognizes in $t_{XX}(0)=\exp(i\wh\fp)$ the shift automorphism. The 
eigenvalues of $\wh\fp$ are the impulsions of the states. Note that 
 the Hamiltonian 
$H_{XX}=\ln(t)'(0)$ (given in (\ref{eq:univXXHam})) vanishes on this subspace. 

There are a priori $\fr$ reference states 
\begin{equation}
\Omega_{a}=\big( e_{a}\big)^{\otimes L}\,,\ a=1,\ldots,\fr=\mbox{rank}\pi
\label{eq:XXvac}
\end{equation}
which have vanishing impulsion and charge 
$L\,\Lambda_{a}$ where $\Lambda_{a}$ is the fundamental weight given in 
(\ref{eq:XXfundw}). 
However, since the algebra 
$\cS=\mbox{End}(\cW)\oplus\mbox{End}(\wb\cW)$ is a symmetry of the 
model, one can restrict itself to highest weight vectors and get the 
remaining states through the action of the step generators of $\cS$.
In fact, in (\ref{eq:XXvac}), there is a unique highest 
weight vector
\begin{equation}
\Omega_{1} = e_{1}\otimes \cdots \otimes e_{1}\,.
\end{equation}
The other states in (\ref{eq:XXvac}) can be obtained through 
iterative action of the symmetry generators $\MM_{a,1}$:
\begin{equation}
\Omega_{a} = \big(\MM_{a,1}\big)^L\ \Omega_{1}
\end{equation}
In the following, we will take  $\Omega_{1}$ is  as the vacuum. 
The other 
states will be  described as excitations above this vacuum, and we 
introduce for $M$ indices $b_{1},\ldots,b_{M}$, and $M$ positions 
$x_{1},\ldots,x_{M}$ the state:
\begin{equation}
\vert\{b\}\,;\,\bx>= 
\underbrace{e_{1}\otimes \cdots \otimes e_{1}}_{x_{1}-1}\otimes {e_{b_{1}}}
\otimes \underbrace{e_{1}\otimes \cdots \otimes e_{1}}_{x_{2}-x_{1}-1}
\otimes {e_{b_{2}}}
\otimes e_{1}\otimes \cdots \otimes e_{1}\otimes {e_{b_{M}}}
\otimes e_{1}\otimes \cdots \otimes e_{1}
\label{eq:omega-bx}
\end{equation}\subsubsection{One excitation states\label{sec:1-excXX}}
We introduce
\begin{eqnarray}
\Phi^1_{a}(p) &=& \sum_{x=1}^L e^{ipx}\,\vert a,x> \qquad 
a=2,\ldots,\mbox{rank}(\pi)=\fr \\
\Phi^1_{\bar a}(p) &=& \sum_{x=1}^L e^{ipx}\,\vert \bar a,x> \qquad 
\bar a=\fr+1,\ldots,\fs =\fr+\wb\fr=\fn+\fm
\end{eqnarray}
where $\vert a,x>$ is defined as in 
(\ref{eq:omega-bx}). The indices $a=2,\ldots,\fr$ correspond to the 
space $\cW$ and the indices $\bar a=\fr+1,\ldots,\fs =\fr+\wb\fr$ 
correspond to the space $\wb\cW$.
Through a direct calculation, it is easy to show that 
\begin{eqnarray}
t(0)\,\Phi^1_{\alpha}(p) &=& e^{ip}\,\Phi^1_{\alpha}(p)\,,\quad 
\alpha=a,\bar a\\ 
H\,\Phi^1_{a}(p) &=& 0 \mb{and} H\,\Phi^1_{\bar a}(p) = 
2\cos(p)\,\Phi^1_{\bar a}(p)
\end{eqnarray}
if $p$ obeys the Bethe ansatz equation (BAE)
\begin{equation}
e^{ipL}=1
\end{equation}
One can gather all these states into a single vector state. The set 
$\{a=2,\ldots,\fr; \bar a=\fr+1,\ldots,\fs =\fr+\wb\fr\}$ is noted 
$\{j=1,\ldots,\fs -1\}$ where the first $\fr-1$ indices are of type `$a$' 
while the $\wb\fr$ last ones are of type `$\bar a$'.
We 
introduce the elementary vectors $u_{j}\in\CC^{\fs -1}$ (with 1 at 
position $j$ and 0 elsewhere): they correspond to the `small' chain of 
the nested Bethe ansatz.  The vector state reads: 
\begin{equation}
\Phi^1(p) = \sum_{j=1}^{\fs -1} \Phi^1_{j+1}(p)\,u_{j} =  
\sum_{x=1}^L e^{ipx}\ \vert x>
\mb{with} \vert x> = \sum_{j=1}^{\fs -1} \vert j,x>\,u_{j}
\end{equation}
Note that in $\vert x>$, $\vert j,x>$ lies on the original `big' chain (of 
length $L$), 
while $u_{j}$ lies on a new `small' chain (here of length 1). The basic idea is to 
`move' the action of the transfer matrix and symmetry generators from 
the `big chain' to the `small one'. 
Indeed, we have
\begin{eqnarray}
t(0)\,\Phi^1(p) &=& e^{ip}\,\Phi^1(p)\\
H\,\Phi^1(p) &=& D(p)\,\Phi^1(p) \mb{with} 
D(p) = 2\cos(p)\,\mbox{diag}\big(\underbrace{0,\ldots,0}_{\fr-1}, 
\underbrace{1,\ldots,1}_{\wb\fr}\big)\qquad
\label{eq:D1}
\end{eqnarray}
The matrix $D(p)$ acts on the small chain (i.e. on the vectors 
$u_{j}$) while $H$ was acting on the big chain (i.e. on the states $\vert 
j,x>$). In the same way,
the charges of the states are given by
\begin{eqnarray}
\MM_{j+1,j+1}\,\Phi^1(p) &=& E_{jj}\,\Phi^1(p)\,,\qquad 
j=1,\ldots,\fs -1 \qquad\\
\MM_{11}\,\Phi^1(p) &=& (L-1)\,\Phi^1(p)
\end{eqnarray}
where $E_{ij}\in\mbox{End}(\CC^{\fs -1})$, $i,j>1$, (the elementary matrix with 1 
at position $(i,j)$ and 0 elsewhere) acts on the small chain. It corresponds to the generator 
of the symmetry generator $\MM_{ij}$ acting on the big chain. 
$\MM_{11}$ (more precisely $L-\MM_{11}$) acts as a scalar and
corresponds to the excitation number.

\subsubsection{Two excitation states and scattering matrix}
We look for eigenstates $\Phi^2_{i,j}(p_{1},p_{2})$ describing two 
excitations of type $i$ and $j$, and with impulsion $p_{1}$ and 
$p_{2}$. We gather these states into a single vector
\begin{equation}
\Phi^2(p_{1},p_{2}) = \sum_{i,j=1}^{\fs -1} 
\Phi^2_{i+1,j+1}(p_{1},p_{2})\,u_{i}\otimes u_{j}
\end{equation}
defining a length 2 `small chain' (carried by the vectors $u_{j}$).
The construction is done in the following way:
\begin{eqnarray}
\Phi^2(p_{1},p_{2}) &=& \sum_{1\leq x_{1}<x_{2}\leq L} 
\Big\{ e^{i\,p\cdot x}\,\II_{\fs -1}\otimes \II_{\fs -1}
+e^{i\,\gamma(p)\cdot x}\,P_{12}\,\cS_{12}(p_{1},p_{2}) \Big\}
\,\vert x_{1},x_{2}>\qquad
\\
\vert x_{1},x_{2}> &=& \sum_{i,j=1}^{\fs -1} \vert i+1,j+1;x_{1},x_{2}>
\,u_{i}\otimes u_{j}\\
\cS_{12}(p_{1},p_{2}) &=& e^{-ip_{1}} \pir\otimes \wb\pi 
+e^{ip_{2}} \wb\pi \otimes \pir -P_{12}\Big(  \pir\otimes \pir+
\wb\pi \otimes \wb\pi \Big) 
\label{eq:SmatXX}
\end{eqnarray}
where we have  introduced $p\cdot x=p_{1}x_{1}+p_{2}x_{2}$ and 
$\gamma(p)\cdot x=p_{2}x_{1}+p_{1}x_{2}$.
$P_{12}$ is the (graded) permutation,
$\wb\pi $ is the projector on $\wb\cW$ and 
$\pir=\II_{\fs -1}-\wb\pi $. 
Since the scattering matrix $\cS$ acts on the small chain, the projector 
$\pir$ is the (lower rank) counter part in the small chain of the 
projector $\pi$ (that acts in the big chain). We kept the same notation 
$\wb\pi$ for both of the projectors $\II_{\fs}-\pi$ and 
$\II_{\fs-1}-\pir$ because they are obviously isomorphic. This 
asymmetric situation is due to our choice of the vacuum, that belongs 
to $\cW=\pi(\cV)$, not to $\wb\cW$.

The scattering matrix obeys 
Yang-Baxter equation and unitarity relation
\begin{eqnarray}
&&\,\cS_{12}(p_{1},p_{2})\,\cS_{13}(p_{1},p_{3})\,\cS_{23}(p_{2},p_{3})
=\,\cS_{23}(p_{2},p_{3})\,\cS_{13}(p_{1},p_{3})\,\cS_{12}(p_{1},p_{2})
,\\
&&\cS_{12}(p_{1},p_{2})\,\cS_{21}(p_{2},p_{1}) = 
\II_{\fs -1}\otimes\II_{\fs -1}\,,\\
&&\cS_{21}(p_{1},p_{2}) = P_{12}\,\cS_{12}(p_{1},p_{2})\,P_{12}\,,
\end{eqnarray}
 while the braided 
$S$-matrix 
$\check\cS_{12}(p_{1},p_{2})=P_{12}\,\cS_{12}(p_{1},p_{2})$ (which 
appears in $\Phi^2(p_{1},p_{2})$) obeys 
braided Yang-Baxer equation and braided unitarity relation:
\begin{eqnarray}
&&\,\check\cS_{23}(p_{1},p_{2})\,\check\cS_{12}(p_{1},p_{3})\,\check\cS_{23}(p_{2},p_{3})
=\,\check\cS_{12}(p_{2},p_{3})\,\check\cS_{23}(p_{1},p_{3})\,\check\cS_{12}(p_{1},p_{2})
,\label{eq:brYBE}\\
&&\check\cS_{12}(p_{1},p_{2})\,\check\cS_{12}(p_{2},p_{1}) = \II_{\fs -1}\otimes\II_{\fs -1}
\label{eq:brUnit}.
\end{eqnarray}

It is easy to show that 
\begin{eqnarray}
t(0)\,\Phi^2(p_{1},p_{2}) &=& e^{i(p_{1}+p_{2})}\,\Phi^2(p_{1},p_{2})\,,\quad 
\\ 
H\,\Phi^2(p_{1},p_{2}) &=& D(p_{1},p_{2})\,\Phi^2(p_{1},p_{2})  
\mb{with} 
D(p_{1},p_{2}) = D(p_{1})\otimes \II_{\fs -1}+ \II_{\fs -1}\otimes D(p_{2})
\nonumber
\end{eqnarray}
if the BAEs
\begin{eqnarray}
e^{ip_{2}L}\,\Phi^2(p_{1},p_{2}) &=& 
\cS_{12}(p_{1},p_{2})\,\Phi^2(p_{1},p_{2}) \\
e^{ip_{1}L}\,\Phi^2(p_{1},p_{2}) &=& 
\cS_{21}(p_{2},p_{1})\,\Phi^2(p_{1},p_{2}) 
\end{eqnarray}
are satisfied. We have used $D(p)$ defined in (\ref{eq:D1}), leading 
to energies 0, $2\cos(p_{1})$, $2\cos(p_{2})$ and $2\cos(p_{1})+2\cos(p_{2})$. 
The charges of the states are given by
\begin{eqnarray}
\MM_{j+1,j+1}\,\Phi^2(p_{1},p_{2}) &=& 
\Big(E_{jj}\otimes\II_{\fs -1}+\II_{\fs -1}\otimes E_{jj}\Big)
\,\Phi^2(p_{1},p_{2})\,,\quad 
j=1,\ldots,\fs -1 \qquad\qquad\\
\MM_{11}\,\Phi^2(p_{1},p_{2}) &=& (L-2)\,\Phi^2(p_{1},p_{2})
\end{eqnarray}
Again, the action of the Hamiltonian $H$ and symmetry generators 
$\MM_{jj}$, $j>1$ have been `moved' to matrices acting on the small 
chain. $L-\MM_{11}$ is the excitation number. Since all the matrices 
are diagonal, we have indeed eigenvectors of the Hamiltonian and 
symmetry generators.

Remark the property
\begin{eqnarray}
&&\Phi^2 (p_{2},p_{1}) = \check S_{12}(p_{1},p_{2})^{-1}\, 
\Phi^2 (p_{1},p_{2})= \check S_{12}(p_{2},p_{1})\, 
\Phi^2 (p_{1},p_{2})
\end{eqnarray}
that ensures that we can impose $p_{1}<p_{2}$. 

The explicit form of the BAE depends on the type of excitation one 
considers. Looking at their projection on vectors 
$u_{i}\otimes u_{j}$ with $i,j<\fr$, one gets the BAE for type $a,b$ 
excitations:
\begin{equation}
e^{ip_{j}L}=\omega\,,\qquad j=1,2\mb{with} \omega^2=1
\end{equation}
If one projects on $u_{i}\otimes u_{j}$ with $i,j\geq\fr$,
one gets the BAE for type $\bar a,\bar b$ 
excitations:
\begin{equation}
e^{ip_{j}L}=\omega\,,\qquad j=1,2
\end{equation}
If one projects on $u_{i}\otimes u_{j}$ with $i<\fr$ and $j\geq\fr$,
one gets the BAE for type $a,\bar a$ 
excitations:
\begin{equation}
e^{ip_{1}(L-1)}=1 \mb{and} e^{i(p_{1}+p_{2})L}=1
\end{equation}
where $p_{1}$ is attached to the type $a$ excitation.

If one projects on $u_{i}\otimes u_{j}$ with $i\geq\fr$ and $j<\fr$,
one gets the BAE for type $\bar a,a$ 
excitations:
\begin{equation}
e^{ip_{2}(L-1)}=1 \mb{and} e^{i(p_{1}+p_{2})L}=1
\end{equation}
where $p_{2}$ is attached to the type $a$ excitation.

\subsubsection{$M$ excitation states and BAEs\label{sec:M-excXX}}
We consider general states , $\Phi^{M}_{\{j\}}(\bp)$, with $M$ excitations 
of momenta $p_{m}$, $m=1,\ldots,M$, gathered into a vector $\bp$. 
If $M'$ is the total number of type `unbarred' excitations, we note
 their corresponding momentum $q_{n}$, $n=1,\ldots,M'$, gathered in a 
 vector $\bq$. In the 
same way, for $M''=M-M'$ the total number of type `barred'
excitations, we noted $\wb q_{n}$, $n=1,\ldots,M''$ their momentum, 
gathered in $\wb\bq$. 
Hence we have 
\beq
\{p_{1},p_{2},\ldots,p_{M}\}=\{q_{1},q_{2},\ldots,q_{M'}\}
\cup\{\wb q_{1},\wb q_{2},\ldots,\wb q_{M''}\}\,.
\eeq
Then, the state
 $\Phi^{M}_{\{j\}}(\bp)$ is characterized by:
\beq
\Phi^{M}_{\{j\}}(\bp)\ :\ \begin{cases}
\mbox{$M$ excitations above the vacuum}\\
\displaystyle\mbox{Momentum: }\ |\bp|=\sum_{m=1}^{M}p_{m}
\equiv|\bq|+|\wb\bq|=\sum_{n=1}^{M'}q_{n}+\sum_{n=1}^{M''}\wb q_{n}\\
\displaystyle\mbox{Charge w.r.t. the symmetry 
algebra: }\ Q=\sum_{m=1}^M \Lambda_{j_{m}}\\
\displaystyle\mbox{Energy: }E=\sum_{n=1}^{M''}2\cos(\wb q_{n})
\end{cases}
\label{eq:dataPhi}
\eeq
 where the weights 
$\Lambda_{j}$ are given in (\ref{eq:XXfundw}). 

All the states with $M$ excitations can be gathered into a single vector
\begin{eqnarray}
\Phi^{M}(\bp) = \sum_{\{j\}} \Phi_{\{j\}}^{M}(\bp)\,u_{j_{1}}\otimes 
u_{j_{2}}\otimes \ldots \otimes u_{j_{M}}\in
\underbrace{\CC^{\fm -1|\fn}\otimes\ldots\otimes\CC^{\fm -1|\fn}}_{M}
\label{eq:def-PhiM}
\end{eqnarray}
describing a small chain of length $M$. 
Then, as for one and two excitation states, the action of the 
different integrals of motion can be `transfered' from the original 
(`big') chain to the new `small' chain:
\begin{eqnarray}
t(0)\,\Phi^{M}(\bp) &=& e^{i\vert\bp\vert}\,\Phi^{M}(\bp)
\label{eq:t0PhiM}\\
H\,\Phi^{M}(\bp) &=& D(\bp)\,\Phi^{M}(\bp) 
\label{eq:HPhiM}\\
D(\bp) &=& \sum_{m=1}^{M} \underbrace{\II_{\fs -1}\otimes \ldots\otimes 
\II_{\fs -1}}_{m-1}\otimes D(p_{m}) \otimes \underbrace{\II_{\fs -1}\otimes \ldots\otimes 
\II_{\fs -1}}_{M-m} \equiv \sum_{m=1}^{M}D_{m}(p_{m})\qquad\\
\MM_{jj}\,\Phi^{M}(\bp) &=& \EE_{j-1,j-1}\,\Phi^{M}(\bp)\,,\ 
j=2,\ldots,\fr+\wb\fr
\label{eq:MkkPhiM}\\
\EE_{jj} &=& \sum_{m=1}^{M} \underbrace{\II_{\fs -1}\otimes \ldots\otimes 
\II_{\fs -1}}_{m-1}\otimes E_{jj} \otimes \underbrace{\II_{\fs -1}\otimes \ldots\otimes 
\II_{\fs -1}}_{M-m} \equiv \sum_{m=1}^{M}E_{jj}^{(m)}
\qquad\\
\MM_{11}\,\Phi^{M}(\bp) &=& (L-M)\,\Phi^{M}(\bp)\,. 
\label{eq:M11PhiM}
\end{eqnarray}
In equalities (\ref{eq:t0PhiM})-(\ref{eq:M11PhiM}), the left-hand 
sides correspond to action on the original chain, while the 
right-hand sides corresponds to their counter-part on the `small' chain.
All the matrices in the r.h.s. are diagonal, and the projection of 
these r.h.s. on a generic state $u_{j_{1}}\otimes 
u_{j_{2}}\otimes \ldots \otimes u_{j_{M}}$ reproduces the data 
(\ref{eq:dataPhi}).

\null
 
The BAEs of the model take the form
\begin{eqnarray}
e^{ip_{j}L}\,\Phi^M(\bp) &=& 
\cS_{j+1,j}\,\cS_{j+2,j}\,\ldots\cS_{Mj}\,
\cS_{1j}\,\cS_{2j}\,\ldots\cS_{j-1,j}\,\Phi^M(\bp) \qquad
 j=1,\ldots,M\qquad
\end{eqnarray}
where $\cS_{jk}\equiv\cS_{jk}(p_{j},p_{k})$ is the two-body 
scattering matrix (\ref{eq:SmatXX}) acting in the spaces $j$ and 
$k$ of the tensor product explicited in (\ref{eq:def-PhiM}).
To compute them explicitly, we introduce the order $\prec$ 
 defined by
\begin{equation}
j+1\prec j+2\prec\ldots\prec M\prec1\prec2\prec\ldots\prec j-1
\end{equation}
Any set of indices $\{j_{1},j_{2},\ldots,j_{n}\}$ ordered 
accordingly, $j_{1}\prec j_{2}\prec \ldots\prec j_{n}$, 
will be noted $[j_{1},j_{2},\ldots,j_{n}]_{\prec}$. 
Then, from the form of $\cS_{12}$, one computes
\begin{eqnarray}
&&\cS_{j+1,j}\,\cS_{j+2,j}\,\ldots\cS_{Mj}\,
\cS_{1j}\,\cS_{2j}\,\ldots\cS_{j-1,j}
\ =\ \sum_{n=0}^M \,\sum_{\{j\}_{n}^{\prec}\oplus\{k\}}\, (-1)^{n}\,
P_{jj_{1}}\,P_{jj_{2}}\ldots P_{jj_{n}}\,\Big\{\nonu
&&\qquad\qquad\ \pir_{j}\,\pir_{j_{1}}\,\ldots \pir_{j_{n}}\,
\wb\pi_{k_{1}}\ldots
\wb\pi_{k_{M-1-n}}\,\exp \Big(i\,(M-1-n)\,p_{j}\Big)
\nonu
&&\qquad\qquad +\,\wb\pi_{j}\,\wb\pi_{j_{1}}\,\ldots 
\wb\pi_{j_{n}}\, 
\pir_{k_{1}}\ldots \pir_{k_{M-1-n}}\,
\exp \Big(-i\sum_{\ell=1}^{M-1-n} p_{k_{\ell}}\Big)\Big\}
\label{eq:prodS}\qquad
\end{eqnarray}
Above, the sum on $\{j\}_{n}^{\prec}\oplus\{k\}$ runs on 
partitions of $[1,M]\setminus\{j\}$, where the first set 
(of cardinality $n$) is ordered according to $\prec$,
$\{j\}_{n}^{\prec}= [j_{1},j_{2},\ldots,j_{n}]_{\prec}$,
 while the second set $\{k\}=\{k_{1},k_{2},\ldots,k_{M-1-n}\}$ is its complementary set.
$\pir_{k}$ is the projector operator $\pir$ in space $k$ (of the `small 
chain').

Applying (\ref{eq:prodS}) to $\Phi^M(\bp)$ leads to the BAEs for the XX model. 
To compute them, we remark that the operator 
$P_{jj_{1}}\,P_{jj_{2}}\ldots P_{jj_{n}}$ corresponds to 
the cyclic permutation of the spaces $j,j_{1},\ldots,j_{n}$ in the 
small chain. Its eigenvalues are $(\omega_{n})^k$, $k=1,2\ldots,n$, 
where $\omega_{n}=e^{2i\pi/n}$. Moreover, this 
operator commutes with diagonal matrix within the brackets of 
eq. (\ref{eq:prodS}) and does not change the type of excitation.
 Thus, the BAEs take the form
\begin{eqnarray}
&&\exp\Big(i\,q_{n}\,(L-M'')\Big) =
(-1)^{M'-1}\,(\omega_{M'})^n\ ,\quad
n = 1,2,...,M'
\\
&&\exp\Big(iL\,\wb q_{n}\Big) = (\omega_{M''})^n\,
\exp\Big(-i\,\vert\bq\vert\Big)\,,\quad
n = 1,2,\ldots,M''\\
&&\mb{with} (\omega_{M'})^{M'}=1
\,,\quad (\omega_{M''})^{M''}=1
\mb{and} \vert \bq\vert=\sum_{n=1}^{M'} q_{n}
\nonumber
\end{eqnarray}
Remark that multiplying together all the BAEs one gets
\begin{equation}
\exp\Big(iL\,\vert\bp\vert\Big)=1\,.
\end{equation}

\section{Universal Hubbard models\label{sect:univHub}}
\label{sec:Hub}

Starting with universal XX models, one can build universal 
Hubbard models, in the same way it has been done for usual and super 
Hubbard models \cite{book,XX}. To simplify the presentation, we present 
the construction in the case of $gl(\fm|\fn)$, but obviously the 
results are valid for any universal Hubbard model.

\subsection{R-matrices} 

\subsubsection{R-matrix for universal Hubbard 
models\label{sec:univR-Hub}} 

We start with the $R$-matrices of two universal
XX models,
$R^{\uparrow}_{12}(\lambda)$ and $R^{\downarrow}_{12}(\lambda)$, 
 leaving in two different sets of  spaces that we label by
$\uparrow$ and $\downarrow$. Let us stress that the two XX models can be based on two
different (graded) vector spaces $\cV_{\uparrow}$ and 
$\cV_{\downarrow}$, with two different projectors
$\pi_{\uparrow}$ and $\pi_{\downarrow}$, associated to the sets
$\cN_{\uparrow}$ and $\cN_{\downarrow}$. 

The Hubbard model is constructed from the coupling of these two XX 
models. Its $R$-matrix reads:
\begin{equation}
R^{\uparrow\downarrow}_{12}(\lambda_{1},\lambda_{2}) =
R^{\uparrow}_{12}(\lambda_{12})\,R^{\downarrow}_{12}(\lambda_{12}) +
\frac{\sin(\lambda_{12})}{\sin(\lambda'_{12})} \,\tanh(h'_{12})\,
R^{\uparrow}_{12}(\lambda'_{12})\,C^{\uparrow}_{1}\,
R^{\downarrow}_{12}(\lambda'_{12})\,C^{\downarrow}_{1}
\label{def:R-XXfus}
\end{equation}
where again $\lambda_{12}=\lambda_{1}-\lambda_{2}$ and 
$\lambda'_{12}=\lambda_{1}+\lambda_{2}$. 
The definition of the parameter
$h'_{12}=h(\lambda_{1})+h(\lambda_{2})$ is given below.
It is easy to show that this 
R-matrix is symmetric
\begin{equation}
R^{\uparrow\downarrow}_{12}(\lambda_{1},\lambda_{2}) =
R^{\downarrow\uparrow}_{21}(\lambda_{1},\lambda_{2}) \,,
\end{equation}
regular
\begin{equation} 
R^{\uparrow\downarrow}_{12}(\lambda_{1},\lambda_{1}) = 
P^{\uparrow\downarrow}_{12}=P^{\uparrow}_{12}\,P^{\downarrow}_{12}
\end{equation}
and obeys the unitarity relation
\begin{eqnarray}
&&\hspace{-2.1ex}R^{\uparrow\downarrow}_{12}(\lambda_{1},\lambda_{2})\,
R^{\uparrow\downarrow}_{21}(\lambda_{2},\lambda_{1}) = 
\left( \cos^{4}(\lambda_{12})
-\Big(\frac{\sin(\lambda_{12})}{\sin(\lambda'_{12})}\,\tanh(h'_{12})\Big)^{2}\right)\,
\II^{\uparrow}_{12}\otimes\II^{\downarrow}_{12}\nonu
&&\mb{where} \II_{12}=\II\otimes\II
\label{unit:R-XXfus}
\end{eqnarray}
\begin{property}
When the function $h(\lambda)$ is given by 
\begin{equation}
\sinh(2h)=U\, \sin(2\lambda) \label{eq:h-lambda}
\end{equation}
for some (free) parameter $U$, the R-matrix (\ref{def:R-XXfus}) obeys YBE:
\begin{eqnarray}
R^{\uparrow\downarrow}_{12}(\lambda_{1},\lambda_{2})\, 
R^{\uparrow\downarrow}_{13}(\lambda_{1},\lambda_{3})\, 
R^{\uparrow\downarrow}_{23}(\lambda_{2},\lambda_{3}) 
&=& 
R^{\uparrow\downarrow}_{23}(\lambda_{2},\lambda_{3})\, 
R^{\uparrow\downarrow}_{13}(\lambda_{1},\lambda_{3})\, 
R^{\uparrow\downarrow}_{12}(\lambda_{1},\lambda_{2})\,.
\end{eqnarray}
In that case, the coefficient in (\ref{unit:R-XXfus}) can be rewritten as
\begin{equation}
\cos^{2}(\lambda_{12})\,\left( \cos^{2}(\lambda_{12})
-\Big(\frac{\tanh(h_{12})}{\cos(\lambda'_{12})}\,\Big)^{2}\right)
\end{equation}
where $h_{12}=h(\lambda_{1})-h(\lambda_{2})$.
\end{property}
\prf 
Again, as remarked in \cite{XX}, the proof relies only on the 
properties (\ref{eq:univpptC}), (\ref{eq:univSig-C}) and follows 
the steps of the original proof by Shiroishi
\cite{shiro}, in the same way it has been done for algebras in 
\cite{book}. Hence, the choice of the projector does not affect it.
Moreover,
it was already noticed in \cite{book} that one can couple
two XX models based on different $gl(\fm)$ algebras: this obviously
extends to general (graded) vector spaces $\cV$.
\finprf

\subsubsection{Gauge version of universal Hubbard models} 
As for the usual Hubbard model, one can introduce a gauged version 
of the above R-matrix. 
It is defined by
\begin{eqnarray}
&&\cR_{12}(\lambda_{1},\lambda_{2}) = 
e^{\half h_{1}\,C^{\uparrow}_{1}C^{\downarrow}_{1}}\,e^{\half
h_{2}\,C^{\uparrow}_{2}C^{\downarrow}_{2}}\,
R^{\uparrow\downarrow}_{12}(\lambda_{1},\lambda_{2}) 
\, e^{-\half h_{1}\,C^{\uparrow}_{1}C^{\downarrow}_{1}}\, e^{-\half
h_{2}\,C^{\uparrow}_{2}C^{\downarrow}_{2}}\nonu
&&\mb{where} h_{j}=h(\lambda_{j})\ ,\quad j=1,2
\label{def:RHub}
\end{eqnarray}
Being a gauged version of the previous R-matrix, 
$\cR_{12}(\lambda_{1},\lambda_{2})$ also obeys YBE,
is unitary and regular. This gauged version is used in usual Hubbard 
model to make contact between the above construction and the 
Hubbard R-matrix as it has been originally built by Shastry.

\subsection{Monodromy matrices, transfer matrices and Hamiltonians} 
We remind that for given vector spaces $\cV_{\downarrow}$ and 
$\cV_{\uparrow}$, the different 
possible projectors $\pi_{\downarrow}$ and $\pi_{\uparrow}$ give different R-matrices
 with, as we shall see, a different symmetry (super)algebra.

We consider the `reduced' monodromy matrix
\begin{equation}
L_{a<b_{1}\ldots b_{L}>}(\lambda)=
\cR_{ab_{1}}(\lambda,0)\ldots \cR_{ab_{L}}(\lambda,0)
\end{equation}
and, when the trace is well-defined, its transfer matrix
$$
t(\lambda)=tr_{a}L_{a<b_{1}\ldots b_{L}>}(\lambda)
$$
Then, one gets
\begin{eqnarray}
 [H, t(\lambda)] = 0\ ,\quad \forall \lambda\ , \mb{for}
H = H(0)=t(0)^{-1}\,t'(0) 
\end{eqnarray}
This `reduced' monodromy matrix is just the one used to define the
Hubbard model; one can compute 
\begin{equation}
\cR_{12}(\lambda,0) = \frac{1}{\cosh(h)}\,
I^{\uparrow\downarrow}_{1}(h)\,R^{\uparrow}_{12}(\lambda)\,R^{\downarrow}_{12}(\lambda)\,I^{\uparrow\downarrow}_{1}(h)
\end{equation}
where
\begin{equation}
I^{\uparrow\downarrow}_{1}(h) = \cosh(\frac{h}{2})\,\II\otimes\II
+\sinh(\frac{h}{2})\,C^{\uparrow}_{1}\,C^{\downarrow}_{1}
\end{equation}

The explicit form of the Hubbard Hamiltonian reads
\begin{equation}
H = \sum_{j=1}^{L}H_{j,j+1}
\label{eq:HubHam}
\end{equation}
with 
\begin{equation}
H_{j,j+1}
=\Sigma^{\uparrow}_{j,j+1}\,P^{\uparrow}_{j,j+1}
+\Sigma^{\downarrow}_{j,j+1}\,P^{\downarrow}_{j,j+1}
+U\,C^{\uparrow}_{j}\,C^{\downarrow}_{j}
\label{eq:twositesham}
\end{equation}
where we have used periodic boundary conditions.
One can see that the kinetics is dictated by the XX models:
barred particles moves `almost freely' with the noticeable exception
that $\bar a^{\uparrow}$
and $\bar b^{\uparrow}$ (or $\bar a^{\downarrow}$
and $\bar b^{\downarrow}$) cannot cross.
Unbarred particles of type up (resp. down) are displaced by barred 
particles of same type. There is interaction only between `up' and 
`down' particles, and the sign of the interaction depends on their `bar' 
or `unbar' type.

\subsection{Symmetries} 
We generalize the results obtained for $su(\fm )$ Hubbard
models (see for instance \cite{maasa2,book}) and $gl(\fm |\fn )$ Hubbard 
models \cite{XX}. 
\begin{proposition}
The transfer matrix of generalized Hubbard  models admits
as symmetry (super)algebra
$$
End(\cV^{\uparrow}_{0})\oplus End(\cV^{\uparrow}_{1})
\oplus End(\cV^{\downarrow}_{0})\oplus End(\cV^{\downarrow}_{1})\,,
$$ 
each of the $End(\cV^{\eps}_{0})\oplus End(\cV^{\eps}_{1})$, 
$\eps=\uparrow,\downarrow$ corresponding to the
symmetry of one XX model.

As a consequence this symmetry is also valid for the 
 Hubbard Hamiltonian.
\end{proposition}
\prf To prove this symmetry, it is
sufficient to remark that
\begin{equation}
\MM\,C=C\,\MM
\end{equation}
where $\MM=\MM^{\uparrow}+\MM^{\downarrow}$ and 
$\MM^\eps\in End(\cV^{\eps}_{0})\oplus End(\cV^{\eps}_{1})$, 
$\eps=\uparrow,\downarrow$. 
Thus, one gets 
\begin{equation}
[R_{12}(\lambda,0)\,,\, \MM^{\uparrow}_{1}+\MM^{\uparrow}_{2}]=0
=[R_{12}(\lambda,0)\,,\,\MM^{\downarrow}_{1}+\MM^{\downarrow}_{2}]
\end{equation}
where $R_{12}(\lambda,0)$ is the $R$-matrix of the universal
Hubbard model.

As far as Hamiltonians and transfer matrices are concerned, the generators of the 
symmetry have the form
\begin{equation}
\MM^{\uparrow}=\sum_{j=1}^{L}\MM^{\uparrow}_{j} \mb{and} 
\MM^{\downarrow} =\sum_{j=1}^{L}\MM^{\downarrow}_{j}
\end{equation}
\finprf
The eigenstates of the transfer matrix will be also eigenstate of the 
Cartan generators $\MM^\eps_{aa}$, 
$a=1,\ldots,dim\cV^\eps=d^\eps$, $\eps=\uparrow,\downarrow$. 
The corresponding charges will be noted 
$\Lambda^\eps=(\lambda_{1}^\eps,\ldots,\lambda^\eps_{d})$.

\section{BAE for universal Hubbard models\label{sect:BAEunivHub}}
We follow the same steps as in section \ref{sec:BAE-XX}.

\subsection{Scattering matrix}
\subsubsection{Pseudo-vacua sector}
The full space of states is now 
$(\cV^\uparrow\otimes\cV^\downarrow)^{\otimes L}$, and we 
 consider the subspace 
$\cW_{vac}=(\cW^\uparrow\otimes\cW^\downarrow)^{\otimes L}$. In this 
 subspace, the Hubbard transfer matrix takes a factorized form:
\begin{equation}
t(\lambda)\Big|_{\cW_{vac}} = \big(1+\tanh(h)\big)^L\, 
t_{XX}^\uparrow(\lambda)\Big|_{\cW_{vac}^\uparrow}
\ t_{XX}^\downarrow(\lambda)\Big|_{\cW_{vac}^\downarrow}
\end{equation}
where $h(\lambda)$ has been given in (\ref{eq:h-lambda}).
The eigenstates of this sector also take a factorized form
\begin{eqnarray}
\Phi^{M\,M'}_{\{ j\},\{ j'\}}(\bp,\bp') &=&
\Phi^{M\,\uparrow}_{\{ j\}}(\bp)\ \Phi^{M'\,\downarrow}_{\{ j'\}}(\bp')
\end{eqnarray}
with eigenvalues
\begin{eqnarray}
&&t(\lambda)\,\Phi^{M\,M'}_{\{ j\},\{ j'\}}(\bp,\bp') 
=\cE(\bp,\bp';\lambda)\ 
\Phi^{M\,M'}_{\{ j\},\{ j'\}}(\bp,\bp')\\
&&\cE(\bp,\bp';\lambda)=\big(1+\tanh(h)\big)^L\ 
\Big( (\cos\lambda)^L\,e^{i\,|\bp|} +(\sin\lambda)^L\,\wb\fr^\uparrow 
\Big)\, \Big( (\cos\lambda)^L\,e^{i\,|\bp'|} 
+(\sin\lambda)^L\,\wb\fr^\downarrow \Big).
\qquad\quad
\end{eqnarray}
Above, we have introduced 
$\wb\fr^\uparrow=\mbox{rank}(\wb\pi^\uparrow)$ and 
$\wb\fr^\downarrow=\mbox{rank}(\wb\pi^\downarrow)$.
The charges of the states read
\ben
\MM^\uparrow\,\Phi^{M\,M'}_{\{ j\},\{ j'\}}(\bp,\bp')=\Lambda^\uparrow\,
\Phi^{M\,M'}_{\{ j\},\{ j'\}}(\bp,\bp') \mb{with}
\Lambda^\uparrow=(L-M)\,\Lambda^\uparrow_{1}
+\sum_{m=1}^{M}\Lambda^\uparrow_{j_{m}}
\\
\MM^\downarrow\,\Phi^{M\,M'}_{\{ j\},\{ j'\}}(\bp,\bp')=\Lambda^\downarrow\,
\Phi^{M\,M'}_{\{ j\},\{ j'\}}(\bp,\bp') \mb{with}
\Lambda^\downarrow=(L-M')\,\Lambda^\downarrow_{1}
+\sum_{m=1}^{M'}\Lambda^\downarrow_{j'_{m}}\,.
\een

Their momentum is given by
\begin{eqnarray}
\wh\fp\,\Phi^{M\,M'}_{\{ j\},\{ j'\}}(\bp,\bp')
&=&i\,\ln\cE(\bp,\bp';0)\,
\Phi^{M\,M'}_{\{ j\},\{ j'\}}(\bp,\bp')
=\Big(\sum_{m=1}^Mp_{m}+\sum_{m=1}^{M'}p'_{m}\Big)\,
\Phi^{M\,M'}_{\{ j\},\{ j'\}}(\,\bp,\bp')\nonu
&\equiv& 
\Big(\vert\bp\vert+\vert\bp'\vert\Big)\,
\Phi^{M\,M'}_{\{ j\},\{ j'\}}(\bp,\bp')
\end{eqnarray}
and their energy reads
\begin{equation}
H\,\Phi^{M\,M'}_{\{ j\},\{ j'\}}(\bp,\bp')
= \frac{\big[\frac{d}{d\lambda}\cE(\bp,\bp';\lambda)\big]_{\lambda=0}}
{\cE(\bp,\bp';0)}\,\Phi^{M\,M'}_{\{ j\},\{ j'\}}(\bp,\bp')
= U\,L\ \Phi^{M\,M'}_{\{ j\},\{ j'\}}(\bp,\bp')
\end{equation}

\subsubsection{General excitations}
Now, we perform general excitations above the vacuum 
$\Omega_{1}^\uparrow\otimes\Omega_{1}^\downarrow$. We note 
\begin{eqnarray}
&&\fs^\eps =\fn^\eps+\fm^\eps=\fr^\eps+\wb\fr^\eps\,,\quad
\fr^\eps=\mbox{rank}(\pi^\eps)\,,\quad 
\wb\fr^\eps=\mbox{rank}(\wb\pi^\eps)\,,\qquad \eps=\uparrow,\downarrow
\\
&&\fs =\fs^\uparrow+\fs^\downarrow\,.
\end{eqnarray}
 We will have four types of excitations:
\beq
\begin{cases} 
\mbox{`unbarred' of type up:}&(a,\uparrow)\equiv j+1\,,\quad 1\leq j\leq \fr^\uparrow-1 \\
\mbox{`barred' of type up:}&(\bar a,\uparrow)\equiv j+1\,,\quad \fr^\uparrow\leq j\leq \fs ^\uparrow-1 \\
\mbox{`unbarred' of type down:}&(a,\downarrow)\equiv j+1-\fs^\uparrow
\,,\quad \fs^\uparrow\leq j\leq \fs^\uparrow+\fr^\downarrow-2 \\
\mbox{`barred' of type down:}&(\bar a,\downarrow)\equiv j+1-\fs^\uparrow\,,\quad 
\fs^\uparrow+\fr^\downarrow-1\leq j\leq \fs -2 
\end{cases}
\label{eq:toto}
\eeq
The set $\{(\bar a,\eps)\}$ corresponds to the space $\wb\cW^\eps$, 
$\eps=\uparrow,\downarrow$. The set $\{ (a,\eps)\}$ corresponds to the space 
$\cW^\eps$ without the index 1 (that is associated to the vacuum): 
in the following, we will note this reduced space $\cWr^\eps$.

\paragraph{One excitation}
For the states with one excitation, one has just to mimick what has 
been done in section \ref{sec:1-excXX}.
\begin{equation}
\begin{array}{l}\displaystyle
\Phi^1_{a,\eps}(p) = \sum_{x=1}^L e^{ipx}\,\vert a,\eps,x> \qquad 
 a=2,\ldots,\mbox{rank}(\pi^\eps)=\fr^\eps 
\\[1.7ex]
\displaystyle
\Phi^1_{\bar a,\eps}(p) = \sum_{x=1}^L e^{ipx}\,\vert \bar 
a,\eps,x> \qquad 
 \bar a=\fr^\eps+1,\ldots,
\fs^\eps=\fr^\eps+\wb\fr^\eps
\end{array}\,,\quad \eps=\uparrow,\downarrow
\end{equation}
Through a direct calculation, it is easy to show that 
\begin{equation}
\begin{array}{l}\displaystyle
t(0)\,\Phi^1_{\alpha,\eps}(p) = e^{ip}\,\Phi^1_{\alpha,\eps}(p)
\,,\quad 
\alpha=a,\bar a\\[1.2ex]
\displaystyle
H\,\Phi^1_{a,\eps}(p) = UL \mb{and} H\,\Phi^1_{\bar a,\eps}(p) = 
\Big(2\cos(p)+U(L-2)\Big)\,\Phi^1_{\bar a,\eps}(p)
\end{array}\,,\quad \eps=\uparrow,\downarrow\,,
\end{equation}
if $p$ obeys the Bethe ansatz equation (BAE)
\begin{equation}
e^{ipL}=1
\end{equation}
Again, one can gather all these states into a single vector state. The 
labelling of excitations is done as explained in (\ref{eq:toto}):
\begin{eqnarray} 
&&\{a=2,\ldots,\fr^\uparrow \ ; \ 
\bar a=\fr^\uparrow+1,\ldots,\fs^\uparrow=\fr^\uparrow+\wb\fr^\uparrow\}
\quad\to\quad
\{j=1,\ldots,\fs^\uparrow-1\}\\
&&\{a=2,\ldots,\fr^\downarrow \ ; \ 
\bar a=\fr^\downarrow+1,\ldots,\fs^\downarrow=\fr^\downarrow+\wb\fr^\downarrow\}
\quad\to\quad
\{j=\fs^\uparrow,\ldots,\fs -2\}\qquad
\end{eqnarray}
 where the first $\fr^\uparrow-1$ indices are of 
type `$a,\uparrow$', the next $\wb\fr^\uparrow$ are of type `$\bar a,\uparrow$',
and so one.
We 
introduce the elementary vectors $u_{j}\in\CC^{\fs -2}$ (with 1 at 
position $j$ and 0 elsewhere) corresponding to the `small' chain of 
the nested Bethe ansatz.  The vector state reads: 
\begin{equation}
\Phi^1(p) = \sum_{j=1}^{\fs -2} \Phi^1_{j+1}(p)\,u_{j} =  
\sum_{x} e^{ipx}\,\vert x>
\mb{with} \vert x> = \sum_{j=1}^{\fs -2} \vert j,x>\,u_{j}
\end{equation}
Note that in $\vert x>$, $\vert j,x>$ lies on the original `big' chain (of 
length $L$), 
while $u_{j}$ lies on a new `small' chain (here of length 1). 
As in section \ref{sec:1-excXX}, we
`move' the action of the transfer matrix and symmetry generators from 
the `big chain' to the `small one'. 
We get
\begin{eqnarray}
t(0)\,\Phi^1(p) &=& e^{ip}\,\Phi^1(p)\\
H\,\Phi^1(p) &=& \cD(p)\,\Phi^1(p)
\ =\ \Big( (2\cos(p)-2U)\, 
\bar D+UL\,\II_{\fs -2}\Big)\,\Phi^1(p)\\
\bar D &=& \mbox{diag}\Big(\underbrace{0,\ldots,0}_{\fr^\uparrow-1}, 
\underbrace{1,\ldots,1\,}_{\wb\fr^\uparrow},
\underbrace{0,\ldots,0}_{\fr^\downarrow-1}, 
\underbrace{1,\ldots,1}_{\wb\fr^\downarrow}\Big)\qquad
\label{eq:D1-hub}
\end{eqnarray}
The matrix $\cD(p)$ acts on the small chain (i.e. on the vectors 
$u_{j}$) while $H$ was acting on the big chain (i.e. on the states $\vert 
j,x>$). In the same way,
the charges of the states are given by
\begin{eqnarray}
\MM_{j+1,j+1}\,\Phi^1(p) &=& E_{jj}\,\Phi^1(p)\,,\qquad 
j=1,\ldots,\fs -2 \qquad\\
\MM_{11}^{\eps}\,\Phi^1(p) &=& (L-D_{\eps})\,\Phi^1(p) \,,\qquad
\eps=\uparrow,\downarrow\\
D_{\uparrow} &=& \mbox{diag}\Big(\underbrace{1,\ldots,1}_{\fs^\uparrow-1}, 
\underbrace{0,\ldots,0}_{\fs^\downarrow-1}\Big)\mb{and}
D_{\downarrow} = \mbox{diag}\Big(\underbrace{0,\ldots,0}_{\fs^\uparrow-1}, 
\underbrace{1,\ldots,1}_{\fs^\downarrow-1}\Big)\qquad
\end{eqnarray}
where $E_{ij}\in\mbox{End}(\CC^{\fs -2})$, $i,j>1$, (the elementary matrix with 1 
at position $(i,j)$ and 0 elsewhere) acts on the small chain. It corresponds to the generator 
of the symmetry generator $\MM_{ij}$ acting on the big chain. 
($L-\MM_{11}^{\eps}$)  
corresponds to the excitation number for $\eps$ particles 
($\eps=\uparrow,\downarrow$).

\paragraph{Two excitations}
For more than one excitation, a new effect appears with respect to 
the XX models: there can be two excitations at the same site 
(provided there are of $\uparrow$ and $\downarrow$ type). To take it
into account, we perform a change of basis on the states and define:
\begin{equation}
\vert x_{1},x_{2}> = \sum_{i=1}^{\fs -2}\sum_{j=1}^{\fs -2} \vert 
{i+1},{j+1};x_{1},x_{2}>\, u_{i}\otimes u_{j}
\end{equation}
with the convention that 
\begin{equation}
\vert {i+1},{j+1};x,x> = \begin{cases}
0 & i,j\leq \fs^\uparrow-1 \\ 
\half\,\vert {i+1},x>\otimes \vert {j+1},x>
& i \leq \fs^\uparrow-1 < j \\
\half\,\vert {i+1},x>\otimes \vert {j+1},x>
& j \leq \fs^\uparrow-1 < i \\
0 & \fs^\uparrow-1<i,j \end{cases}
\end{equation}
Then, the eigenstates are gathered into a vector
\begin{equation}
\Phi^2(p_{1},p_{2}) = \sum_{i=1}^{\fs -2}\sum_{j=1}^{\fs -2} 
\Phi^2_{i,j}(p_{1},p_{2})\,u_{i}\otimes u_{j}
\end{equation}
We have
\begin{equation}
\Phi^2(p_{1},p_{2}) = \sum_{1\leq x_{1}\leq x_{2}\leq L} \Big\{
e^{i\,p\cdot x}\,\II_{\fs -2}\otimes\II_{\fs -2} + 
e^{i\,\gamma(p)\cdot x}\,P_{12}\,\cS_{12}(p_{1},p_{2}) \Big\}\,
\vert x_{1},x_{2}>
\end{equation}
The scattering matrix is given by
\begin{eqnarray}
\cS_{12}(p_{1},p_{2}) &=& \cS^{X\uparrow}_{12}(p_{1},p_{2})
+\cS^{X\downarrow}_{12}(p_{1},p_{2})+\cS^{\updownarrow}_{12}(p_{1},p_{2})
+\cS^{H}_{12}(p_{1},p_{2})\\[1.2ex]
\cS^{X\eps}_{12}(p_{1},p_{2}) &=& 
e^{-ip_{1}}\,\pir^{\eps}\otimes \wb\pi^{\eps}
+e^{ip_{2}}\,\wb\pi^{\eps}\otimes \pir^{\eps}
-P_{12}\Big(
\pir^{\eps}\otimes \pir^{\eps} +\wb\pi^{\eps}\otimes \wb\pi^{\eps}\Big)
\,,\ \eps=\uparrow,\downarrow\qquad\\
\cS^\updownarrow_{12}&=& 
\pir^{\uparrow}\otimes (\pir^{\downarrow}+\wb\pi^{\downarrow})+
(\pir^{\downarrow}+\wb\pi^{\downarrow})\otimes \pir^{\uparrow}+
\pir^{\downarrow}\otimes \wb\pi^{\uparrow}
+\wb\pi^{\uparrow}\otimes \pir^{\downarrow}
\\
\cS^{H}_{12}(p_{1},p_{2}) &=& \Big(T(p_{1},p_{2})\,\II_{\fs -2}\otimes\II_{\fs -2}
+R(p_{1},p_{2})\,P_{12}\Big)\,
\Big(\wb\pi^{\uparrow}\otimes \wb\pi^{\downarrow}
+\wb\pi^{\downarrow}\otimes \wb\pi^{\uparrow} \Big)
\\
T(p_{1},p_{2}) &=& 
\frac{\sin(p_{1})-\sin(p_{2})}{\sin(p_{1})-\sin(p_{2})-2iU}\\
R(p_{1},p_{2}) &=& 
\frac{2iU}{\sin(p_{1})-\sin(p_{2})-2iU} \ =\ T(p_{1},p_{2}) -1
\end{eqnarray}
where $\pir^{\eps}$ (resp. $\wb\pi^{\eps}$) is the projector on 
$\cWr^\eps$ (resp. $\wb\cW^\eps$), $\eps=\uparrow,\downarrow$.

One recognizes in $\cS^{X\eps}_{12}(p_{1},p_{2})$ the  
scattering matrix of an XX model in the `$\eps$ subsector' 
($\eps=\uparrow,\downarrow$). They 
correspond to the only part of $\cS$ which acts non trivially in the 
$\uparrow\uparrow$ and $\downarrow\downarrow$ sectors. The remaining 
part (acting in the $\downarrow\uparrow$ and $\uparrow\downarrow$ 
sectors) have been divided into a part acting only in the `bar 
sector' (the $\cS^{H}$ matrix, of Heisenberg type) and the rest (the $\cS^\updownarrow$ 
matrix).
\paragraph{Comparison with usual Hubbard model:} the parts 
$\cS^{X\eps}_{12}(p_{1},p_{2})$ and $\cS^{H}_{12}(p_{1},p_{2})$ in 
the scattering matrix are just generalizations of the Hubbard 
scattering matrix to higher dimensional 
case. Note however the projections appearing in these scattering matrices, 
that are 
new w.r.t. the usual Hubbard model: we will comment on this point in section 
\ref{sec:barHub}. The part $\cS^\updownarrow$ is completely new: it 
introduces new physical effects that were not seen in Hubbard, due to 
the `small size' of its vector space.

\subsection{BAEs: a first account}
Once the scattering matrix of the universal Hubbard model is known, 
the technique to obtain the transfer matrix eigenvalues and 
the BAEs is a priori known, see section \ref{sec:Hub-Mexc} below. 
However, if the eigenvalues are easy to deduce, the determination of
the precise form of the BAEs is a more delicate problem. Here, we 
compute them for some subsectors of the theory, leaving the 
determination of their complete form for a further publication. 

\subsubsection{$M$ excitations\label{sec:Hub-Mexc}}
We consider a general state with $M$ excitations, that divides into 
$M^\uparrow$ excitations of type $a$ in the `$\uparrow$ sector', 
$\bar M^\uparrow$ excitations of type $\bar a$ in the `$\uparrow$ sector', 
$M^\downarrow$ excitations of type $a$ in the `$\downarrow$ sector', 
and
$\bar M^\downarrow$ excitations of type $\bar a$ in the `$\downarrow$ sector'.
The construction follows the line of 
section \ref{sec:M-excXX}, with the noticeable exception that there 
can be $\uparrow$ and $\downarrow$ excitations at the same site. 
To take this fact into account, we introduce:
\begin{eqnarray}
\vert \{j\};\bx> &=& \begin{cases}
\otimes_{m=1}^M \vert j_{m};x_{m}> \mb{if all $x_{m}$'s are different}
\\[1.2ex]
0 \mb{if at least three $x_{m}$'s are equal}
\end{cases}\\[1.2ex]
\vert \{j\};\bx>\Big\vert_{x_{m}=x_{m'}} &=& \begin{cases}
0 & \mbox{if }{j},{j'}\leq \fs ^\uparrow-1 \\[1.2ex] 
\half\,\otimes_{m=1}^M \vert j_{m};x_{m}> 
& \mbox{if }{j} \leq \fs ^\uparrow-1 < {j'}\qquad \\[1.2ex]
\half\,\otimes_{m=1}^M \vert j_{m};x_{m}> 
& \mbox{if }{j'} \leq \fs ^\uparrow-1 < {j} \\[1.2ex]
0 &\mbox{if } \fs ^\uparrow-1<{j},{j'} \end{cases}
\end{eqnarray}

Then, the BAEs take the form
\begin{eqnarray}
e^{ip_{j}L}\,\Phi^M(\bp) &=& 
\cS_{j+1,j}\,\cS_{j+2,j}\,\ldots\cS_{Mj}\,
\cS_{1j}\,\cS_{2j}\,\ldots\cS_{j-1,j}\,\Phi^M(\bp) \qquad
 j=1,\ldots,M\qquad
\end{eqnarray}

In the following, we examine the BAEs in 
subsectors that are related to different types of excitations: the 
two XX-type subsectors, where excitations are only of 
type $\uparrow$ or only of type $\downarrow$ ; the `unbarred 
subsector', where excitations are only of type `unbarred' ($\uparrow$ or 
$\downarrow$), and the Hubbard-type 
subsector, where excitations are only of type `bar' ($\uparrow$ or 
$\downarrow$). 

\subsubsection{BAEs for the XX-type subsector}
We introduce the projectors on the `$\uparrow$ sector' and `$\downarrow$ sector'
\begin{equation}
\Pi^{\eps}=\pir^{\eps}+\wb\pi^{\eps}\,,\quad\eps=\uparrow,\downarrow\,.
\end{equation}
It is easy to see that 
\begin{equation}
\Pi^{\eps}\otimes \Pi^{\eps}\,\cS_{12}(p_{1},p_{2}) = \cS^{X\eps}_{12}(p_{1},p_{2})\,
\Pi^{\eps}\otimes \Pi^{\eps}
\qquad \eps=\uparrow,\downarrow
\end{equation}
so that multiplying the BAEs from the left, by $(\Pi^{\eps})^{\otimes 
M}$, one recovers 
 the BAEs of the $XX$ models:
\begin{eqnarray}
e^{ip_{j}L}\,\Phi^M_{\uparrow}(\bp) &=& 
\cS^{X\uparrow}_{j+1,j}\,\cS^{X\uparrow}_{j+2,j}\,\ldots\cS^{X\uparrow}_{Mj}\,
\cS^{X\uparrow}_{1j}\,\cS^{X\uparrow}_{2j}\,\ldots\cS^{X\uparrow}_{j-1,j}\,\Phi^M_{\uparrow}(\bp) 
\nonu
&& j=1,\ldots,M\mb{;}M^\downarrow=\bar M^\downarrow=0
\\[1.7ex]
e^{ip_{j}L}\,\Phi^M_{\downarrow}(\bp) &=& 
\cS^{X\downarrow}_{j+1,j}\,\cS^{X\downarrow}_{j+2,j}\,\ldots\cS^{X\downarrow}_{Mj}\,
\cS^{X\downarrow}_{1j}\,\cS^{X\downarrow}_{2j}\,\ldots\cS^{X\downarrow}_{j-1,j}\,\Phi^M_{\uparrow}(\bp) 
\nonu
&& j=1,\ldots,M\mb{;}M^\uparrow=\bar M^\uparrow=0
\\[1.7ex]
\Phi^M_{\eps}(\bp) &=& 
(\underbrace{\Pi^{\eps}\otimes\Pi^{\eps}\otimes\ldots
\otimes\Pi^{\eps}}_M)\,\Phi^{M}(\bp)\,,\qquad
\eps=\uparrow,\downarrow
\end{eqnarray}
These BAEs corresponds to subsectors where excitations of only 
$\uparrow$ or only $\downarrow$ types are considered. They are of the 
same form that the XX models:
\begin{eqnarray}
&&\exp\Big(i\,q_{n}\,(L-M'')\Big) = 
(-1)^{M'-1}\,(\omega_{M'})^n
\,,\quad n=1,...,M'
\mb{with} (\omega_{M'})^{M'}=1
\\
&&\exp\Big(iL\,\wb q_{n}\Big) = (\omega_{M''})^n\,
\exp\Big(-i\,\vert\bq\vert\Big)
\,,\quad n=1,...,M''
\mb{with} (\omega_{M''})^{M''}=1\,.
\qquad
\end{eqnarray}

\subsubsection{BAEs for the `unbarred sector'}
We consider the `unbarred subsector', i.e. states with unbarred
 excitations (of type $\uparrow$ or $\downarrow$) only. 
The corresponding projector is
\begin{equation}
\Pi = \pir^{\uparrow}+\pir^{\downarrow}\mb{;} \Pi_{12}=\Pi\otimes\Pi
\mb{;} \Pi_{1\ldots M}=
\underbrace{\Pi\otimes\Pi\otimes\ldots\otimes\Pi\otimes\Pi}_{M}
\end{equation}
From the property
\begin{eqnarray}
\Pi_{1\ldots M}\,S_{12}(p_{1},p_{2}) &=& 
S^{un}_{12}(p_{1},p_{2})\, \Pi_{1\ldots M}\\
S^{un}_{12}(p_{1},p_{2})&=& \sum_{\eps=\uparrow,\downarrow}
\left\{\pir^{\eps}\otimes  \pir^{-\eps}-P_{12}\Big(
\pir^{\eps}\otimes \pir^{\eps} \Big)\right\}
\end{eqnarray}
it is easy to see that the calculation is very similar to the XX 
case, with $\pir^{\uparrow},\pir^{\downarrow}$ playing the role of 
$\pir,\wb\pi$ of section \ref{sec:M-excXX}. Using the same notations, one gets
\begin{eqnarray*}
&&\cS^{un}_{j+1,j}\,\cS^{un}_{j+2,j}\,\ldots\cS^{un}_{Mj}\,
\cS^{un}_{1j}\,\cS^{un}_{2j}\,\ldots\cS^{un}_{j-1,j}
\ =\ \sum_{n=0}^M\ \sum_{\{j\}_{n}^{\prec}\oplus\{k\}} (-1)^{n}\,
P_{jj_{1}}\,P_{jj_{2}}\ldots P_{jj_{n}}\,\Big\{
\qquad\nonu
&&
\qquad\qquad 
\pir^{\uparrow}_{j}\,\pir^{\uparrow}_{j_{1}}\,\ldots \pir^{\uparrow}_{j_{n}}\,
\pir^{\downarrow}_{k_{1}}\ldots
\pir^{\downarrow}_{k_{M-1-n}}\,
+\,\pir^{\downarrow}_{j}\,\pir^{\downarrow}_{j_{1}}\,\ldots 
\pir^{\downarrow}_{j_{n}}\, 
\pir^{\uparrow}_{k_{1}}\ldots \pir^{\uparrow}_{k_{M-1-n}}\,\Big\}
\label{eq:prodSunbar}
\end{eqnarray*}
Since in this sector the BAEs take the form
\begin{eqnarray*}
e^{ip_{j}L}\,\Phi^M_{un}(\bp) &=&\cS^{un}_{j+1,j}\,\cS^{un}_{j+2,j}\,\ldots\cS^{un}_{Mj}\,
\cS^{un}_{1j}\,\cS^{un}_{2j}\,\ldots\cS^{un}_{j-1,j}
\ \Phi^M_{un}(\bp) \\
\Phi^M_{un}(\bp) &=& \Pi_{1\ldots M}\,\Phi^M(\bp)
\end{eqnarray*}
we obtain
\begin{eqnarray}
\exp\Big(i\,q_{n}\,L\Big) &=& 
(-1)^{M^{\uparrow}-1}\,(\omega_{M^{\uparrow}})^n
\,,\quad n = 1,2,\ldots,M^{\uparrow}
\mb{with} (\omega_{M^{\uparrow}})^{M^{\uparrow}}=1
\,,\quad\label{eq:BAEunbup}
\\
\exp\Big(i\,q'_{n}\,L\Big) &=& 
(-1)^{M^{\downarrow}-1}\,(\omega_{M^{\downarrow}})^n
\,,\quad n = 1,2,\ldots,M^{\downarrow}
\mb{with} (\omega_{M^{\downarrow}})^{M^{\downarrow}}=1
\label{eq:BAEunbdwn}
\end{eqnarray}
where $M^{\uparrow}$ is the number of $\uparrow$ excitations, and
$M^{\downarrow}=M-M^{\uparrow}$ is the number of $\downarrow$ 
excitations. We noted $q_{n}$, $n=1,2,\ldots,M^{\uparrow}$, the 
momenta of the $\uparrow$ excitations and 
$q'_{n}$, $n=1,2,\ldots,M^{\downarrow}$, the 
momenta of the $\downarrow$ excitations.

Remark that the two series of BAEs (\ref{eq:BAEunbup}) and 
(\ref{eq:BAEunbdwn}) are 
decoupled, and correspond to the `unbarred sector' of each of the 
underlying XX models. They can be obtained separately using the 
projectors 
\begin{equation}
\Pi_{1\ldots M}^\uparrow=
\underbrace{\pir^{\uparrow}\otimes\pir^{\uparrow}\otimes\ldots\otimes\pir^{\uparrow}
\otimes\pir^{\uparrow}}_{M}
\mb{or}
\Pi_{1\ldots M}^\downarrow=
\underbrace{\pir^{\downarrow}\otimes\pir^{\downarrow}\otimes\ldots\otimes\pir^{\downarrow}
\otimes\pir^{\downarrow}}_{M}\,,
\end{equation}
but the present calculation shows that they are complete in this 
subsector.
\subsubsection{The `bar subsector'\label{sec:barHub}}
Following the same lines as in the previous sections, one can
 consider the `bar subsector', i.e. states with $\bar a_{\uparrow}$ and 
$\bar a_{\downarrow}$ excitations only. The corresponding projector is
\begin{equation}
\bar\Pi = \wb\pi^{\uparrow} + \wb\pi^{\downarrow}
\mb{and} \bar\Pi_{1\ldots M}= 
\underbrace{\bar\Pi\otimes\bar\Pi\otimes\ldots\otimes\bar\Pi}_{M}
\,.
\label{eq:Pibar}
\end{equation}
In that case, one has
\begin{eqnarray}
\bar\Pi_{1\ldots M}\,\cS_{12}(p_{1},p_{2}) &=& \cS_{12}(p_{1},p_{2})\,
\bar\Pi_{1\ldots M} \equiv \wb\cS_{12}(p_{1},p_{2})\,\bar\Pi_{1\ldots M}
\nonu
\wb\cS_{12}(p_{1},p_{2})&=& \Big(T(p_{1},p_{2})\,\II_{\fs 
-2}\otimes\II_{\fs -2}
+R(p_{1},p_{2})\,P_{12}\Big)\,
\Big(\wb\pi^{\uparrow}\otimes \wb\pi^{\downarrow}
+\wb\pi^{\downarrow}\otimes \wb\pi^{\uparrow} \Big)
\nonu
&&-P_{12}\,\Big(\wb\pi^{\uparrow}\otimes\wb\pi^{\uparrow}+\wb\pi^{\downarrow}\otimes 
\wb\pi^{\downarrow}\Big)
\end{eqnarray}
One could be tempted to reccognize in $\wb\cS_{12}$, 
the scattering matrix of a generalized XXX model. Indeed,
 specifying the $\uparrow$ or 
$\downarrow$ type only (whatever the indices $\bar a,\bar b,\bar c,\ldots$ 
are), one gets on a state with two excitations:
\begin{eqnarray}
\wb\cS_{12}(p_{1},p_{2})\,\vert{\uparrow\uparrow'}>
&=& -\vert{\uparrow'\uparrow}>
\label{eq:bbupup}\\
\wb\cS_{12}(p_{1},p_{2})\,\vert{\uparrow\downarrow}>
&=& T(p_{1},p_{2})\,\vert{\uparrow\downarrow}>
+R(p_{1},p_{2})\,\vert{\downarrow\uparrow}>
\\
\wb\cS_{12}(p_{1},p_{2})\,\vert{\downarrow\uparrow}>
&=& R(p_{1},p_{2})\,\vert{\uparrow\downarrow}>
+T(p_{1},p_{2})\,\vert{\downarrow\uparrow}>
\\
\wb\cS_{12}(p_{1},p_{2})\,\vert{\downarrow\downarrow'}>
&=& -\vert{\downarrow'\downarrow}>
\label{eq:bbdodo}
\end{eqnarray}
In the case of Hubbard model, where there is only one type of 
$\uparrow$ excitation and one type of $\downarrow$ excitation, one has 
exactly the scattering matrix of the XXX models. This allowed the 
calculation of BAEs of the Hubbard model. 
\underline{However}, if there is more than one type of $\uparrow$ 
or $\downarrow$ excitation, this is not the case anymore. For 
instance, for two types of $\uparrow$ excitations (say $\bar a$ and 
$\bar b$), eq. (\ref{eq:bbupup}) corresponds to
\begin{eqnarray}
\wb\cS_{12}(p_{1},p_{2})\,\vert{\bar a\uparrow;\bar a\uparrow'}>
&=& -\vert{\bar a\uparrow';\bar a\uparrow}>\\
\wb\cS_{12}(p_{1},p_{2})\,\vert{\bar a\uparrow;\bar b\uparrow'}>
&=& -\vert{\bar b\uparrow';\bar a\uparrow}>\\
\wb\cS_{12}(p_{1},p_{2})\,\vert{\bar b\uparrow;\bar a\uparrow'}>
&=& -\vert{\bar a\uparrow';\bar b\uparrow}>\\
\wb\cS_{12}(p_{1},p_{2})\,\vert{\bar b\uparrow;\bar b\uparrow'}>
&=& -\vert{\bar b\uparrow';\bar b\uparrow}>
\een
while a generalized XXX model would act as
\begin{eqnarray}
\wb\cS_{12}(p_{1},p_{2})\,\vert{\bar a\uparrow;\bar a\uparrow'}>
&=& -\vert{\bar a\uparrow';\bar a\uparrow}>\\
\wb\cS_{12}(p_{1},p_{2})\,\vert{\bar a\uparrow;\bar b\uparrow}>
&=& T(p_{1},p_{2})\,\vert{\bar a\uparrow;\bar b\uparrow}>
+R(p_{1},p_{2})\,\vert{\bar b\uparrow;\bar a\uparrow}>
\\
\wb\cS_{12}(p_{1},p_{2})\,\vert{\bar b\uparrow;\bar a\uparrow}>
&=& T(p_{1},p_{2})\,\vert{\bar b\uparrow;\bar a\uparrow}>
+R(p_{1},p_{2})\,\vert{\bar a\uparrow;\bar b\uparrow}>\\
\wb\cS_{12}(p_{1},p_{2})\,\vert{\bar b\uparrow;\bar b\uparrow'}>
&=& -\vert{\bar b\uparrow';\bar b\uparrow}>
\een
This difference just prevents to perform a nesting in the same way it 
is done for Hubbard. Note that $\wb\cS_{12}$ still obeys
\beq
\wb\cS_{12}(p,p)=-P_{12}
\eeq
so that one can define an integrable spin chain associated to the 
nesting in the usual way. However, the exact form of the BAEs for 
this new chain is not known yet. The same is true for the general BAEs 
of the universal Hubbard model. We will come back on this point in a 
further work \cite{vik}.

\section{Perturbative expansion of the Hubbard-like Hamiltonian\label{KSexpansion}}
\newcommand{\site}[1]{\displaystyle\mathop{\otimes}_{#1}}
We expand the Hamiltonian (\ref{eq:HubHam}) and (\ref{eq:twositesham}) in the 
inverse coupling $\frac1{U}$. That expansion has been used 
in \cite{Rej:2005qt} to match the $SU(2)$ dilatation operator with the 
effective Hamiltonian of the Hubbard model. The system was taken at 
half-filling to guarantee the required spin chain behaviour.

Being ultralocal, the potential term $U\sum_j C^{\uparrow}_jC_j^{\downarrow}$ 
is separately diagonalizable on each site with eigenvalues $\pm U$. Indeed, 
they are obtained from the property $C^2=\II$ (\ref{eq:opC}). 

The ground state has eigenvalue $-L U$ and can be obtained if the condition
$C^{\uparrow}_jC_j^{\downarrow}=-1$ is realised on each site. This is
equivalent to demand eigenvalue 1 on each site for the (one-site) projector
\begin{equation}
\frac{1-C^{\uparrow}_jC_j^{\downarrow}}{2}=
\left(\frac{1-C^{\uparrow}_jC_j^{\downarrow}}{2}\right)^2
=\pi^{\uparrow}_j+\pi_j^{\downarrow}-2\pi^{\uparrow}_j \pi_j^{\downarrow}
=\big(\pi^{\uparrow}_j-\pi_j^{\downarrow}\big)^2 
\end{equation}
or the global projector
\begin{equation}
\Pi_{0}=\prod _j  \big(\pi^{\uparrow}_j-\pi_j^{\downarrow}\big)^2 = 
\prod _j  \big(\wb{\pi}^{\uparrow}_j-\wb{\pi}_j^{\downarrow}\big)^2=
\Pi_{0}^2\,.\label{eq:Pi0}
\end{equation}
We observe that it projects on the subspace where, on each site, 
one and only one projector among 
$\wb\pi^{\uparrow}_j \,,\ \wb\pi_j^{\downarrow}$ has nonzero action.
This means that only the following subspaces survive
\begin{equation}\label{ground}
\Pi_0\,:\qquad \cW^{\uparrow} \site j \wb\cW^{\downarrow} \qquad 
\text{or} \qquad \wb\cW^{\uparrow} \site j \cW^{\downarrow}
\end{equation}
Namely we demand that on each site there is one barred particle: double or
empty occupancies of barred particles are prohibited. This may be possible
only  if
the system has precisely $L$ barred particles out of the $2L$ permitted ones. We
say that the system is \textit{half-filled} and we will assume this
condition to perform the perturbative calculations.

It is useful to compare with the ordinary Hubbard model, where
the algebra is realised in terms of fermionic oscillators $c_{\sigma,j}$,
$\,c_{\sigma,j}^{\dagger}$, satisfying
$\{c_{\sigma,j},\,c_{\sigma,j}^{\dagger}\}=1$ $(\sigma=
\uparrow\,,\downarrow)$. There, the projector $\wb\pi^{\sigma}_j$ is equal
to the number operator $n_{\sigma,j}=c_{\sigma,j}^{\dagger}c_{\sigma,j}$, so
in the present general formalism, a vector of $\wb\cW^{\sigma}$ corresponds
to an electron and a vector of $\cW^{\sigma}$ corresponds to a vacancy (in
the Hubbard model we have $\pi^{\sigma}_j=1-n_{\sigma,j}$).

We follow the method introduced by Klein and Seitz \cite{klstz}.
With reference to the Hamiltonian (\ref{eq:twositesham}) and in complete 
analogy with 
previous cases \cite{XX}, we define a hopping operator by
\begin{eqnarray}\label{hopping}
X_{ij}&=&P_{ij}^{\uparrow}\pi_i^{\uparrow}\wb\pi_j^{\uparrow}+
P_{ij}^{\downarrow}\pi_i^{\downarrow}\wb\pi_j^{\downarrow} 
\ =\ \wb\pi_i^{\uparrow}\pi_j^{\uparrow} P_{ij}^{\uparrow}
\pi_i^{\uparrow}\wb\pi_j^{\uparrow}+
\wb\pi_i^{\downarrow}\pi_j^{\downarrow}P_{ij}^{\downarrow}
\pi_i^{\downarrow}\wb\pi_j^{\downarrow}\nonumber
\end{eqnarray}
Intuitively we associate its action to the move of a barred particle from
site $j$ to site $i$. We define hermitian conjugation as
super-transposition (because our operators are real) so that we have
\begin{equation}
X_{ij}^{\dagger}=X_{ji}
\end{equation}
The two-sites Hamiltonian (\ref{eq:twositesham}) takes the form 
\begin{equation}
H_{j,j+1}=X_{j,j+1}+X_{j+1,j}+U\,C^{\uparrow}_{j}\,C^{\downarrow}_{j}
\end{equation}
and is obviously self-adjoint. The perturbing term is 
\begin{equation}\label{pertur}
T=\sum_j (X_{j,j+1}+X_{j+1,j})\,.
\end{equation}
The action of $X_{ij}$ on the vector spaces defined in (\ref{def:univpi})
is easily described by observing the projectors in the second line of 
(\ref{hopping}).
We initially focus on the effect on site $i$:
\begin{equation}\label{action}
\begin{array}{c@{)\qquad}c@{\quad\xrightarrow[]{~X_{ij}~}\quad}c}
1&\cW^{\uparrow} \site i \cW^{\downarrow} & \wb\cW^{\uparrow} \site i 
\cW^{\downarrow}~ \oplus~ \cW^{\uparrow} \site i\wb\cW^{\downarrow}\\
2&\wb\cW^{\uparrow} \site i \wb\cW^{\downarrow} & 0 \\
3&\cW^{\uparrow} \site i \wb\cW^{\downarrow} & \wb\cW^{\uparrow} \site i 
\wb\cW^{\downarrow} \\
4&\wb\cW^{\uparrow} \site i \cW^{\downarrow} & \wb\cW^{\uparrow} \site i 
\wb\cW^{\downarrow} 
\end{array}
\end{equation}
On site $j$ the description is the complementary one, namely:
\begin{equation}\label{actionc}
\begin{array}{c@{)\qquad}c@{\quad\xrightarrow[]{~X_{ij}~}\quad}c}
1&\cW^{\uparrow} \site j \cW^{\downarrow} & 0\\
2&\wb\cW^{\uparrow} \site j \wb\cW^{\downarrow} & \cW^{\uparrow} \site j
\wb\cW^{\downarrow} ~ \oplus~ \wb\cW^{\uparrow} \site j \cW^{\downarrow} \\
3&\cW^{\uparrow} \site j \wb\cW^{\downarrow} & \cW^{\uparrow} \site j 
\cW^{\downarrow} \\
4&\wb\cW^{\uparrow} \site j \cW^{\downarrow} & \cW^{\uparrow} \site j 
\cW^{\downarrow} 
\end{array}
\end{equation}
It is clear that the domain and the codomain of $X_{ij}$ are always disjoint and 
only a ``two-fold'' action can make them the same. We state this as a theorem.
\begin{theorem}\label{oddeven}
Given a site $i$ and an initial configuration choosen among the four listed in 
(\ref{action}), the product of an odd number of operators 
$X_{ij}$ or $X_{ji}$ acting on $i$ with possibly different $j$ 
cannot return to the same initial configuration.
\end{theorem}
\prf The proof is a trivial application of the rules in (\ref{action} and 
\ref{actionc})
\finprf

\noindent
A number of corollaries follow from it and from the action of the projector 
$\Pi_0$:
\begin{gather}
X_{ij}^2\ \Pi_0=0 \label{evenex}\\
(1-\Pi_0)\ X_{ij}\ X_{ji}\ \Pi_0=0 \nonumber\\
X_{j-1,j}\ X_{j+1,j}\  \Pi_0=0 \nonumber\\
\Pi_0\ T^n\ \Pi_0=0 \qquad \text{if }\ n=\text{odd \ and \ } L>n \nonumber
\end{gather}
The condition $L>n$ in (\ref{evenex}) is extremely important. 
If it is removed, new terms known as demi-wrapping $L=n$ and wrapping 
$L<n$ behaviours occur because the perturbative interaction (\ref{pertur}) 
circulates all around the periodic chain. These terms will not be evaluated here.

According to Klein and Seitz \cite{klstz}, the effective Hamiltonian
for the states originated from $\Pi_0$ is
\begin{equation}\label{ks}
H_{\text{eff}}=\frac1{U} H_{\text{eff}}^{(2)}+\frac1{U^3} 
H_{\text{eff}}^{(4)}+ \ldots
\end{equation}
where
\begin{gather}
S=(1-\Pi_0)\frac1{E_0-H_0}(1-\Pi_0)\\
H_{\text{eff}}^{(2)}=\Pi_0 T S T \Pi_0=-\frac14  \Pi_0\ T^2 \ \Pi_0 \\
H_{\text{eff}}^{(4)}=\frac1{16} \Pi_0 T^2 S T^2 \Pi_0+\frac{1}{64}\Pi_0 T^2 \Pi_0 
T^2 \Pi_0
\end{gather}
These expressions can be worked out following Klein and Seitz \cite{klstz},
with a long but simple calculation that is not shown here, using 
the properties (\ref{evenex}).

We point out another property that allows to simplify the expressions:
the projector $\Pi_0$ makes redundant one among the barred and the non-barred projectors.
Indeed, we can write the identity on site $j$ as $\pi^{\sigma}_j+ \wb\pi^{\sigma}_j=1$,
therefore
\begin{equation}
\pi^{\sigma}_j\Pi_0=\pi^{\sigma}_j(\pi^{-\sigma}_j+ \wb\pi^{-\sigma}_j)\Pi_0=
\pi^{\sigma}_j \wb\pi^{-\sigma}_j\Pi_0=\wb\pi^{-\sigma}_j\Pi_0
\end{equation}
and we can write all the expressions using only the non-barred projectors.

\subsection{Second order Hamiltonian}
A direct calculation shows that for $L>2$ the second order effective 
Hamiltonian is
\begin{eqnarray}\label{secondord}
H_{\text{eff}}^{(2)}&=&\sum _j H_{\text{eff}\ j,j+1}^{(2)} \\
&=&\Pi_0 \Big(  2\sum _j \big(1+P^{\uparrow}_{j,j+1} P^{\downarrow}_{j,j+1}\big) 
\big(\pi^{\uparrow}_j\,\pi^{\downarrow}_{j+1}+
\pi_j^{\downarrow}\,\pi^{\uparrow}_{j+1} \big) \Big)  \Pi_0 \nonumber
\end{eqnarray}
The structure of the two-sites Hamiltonian $H_{\text{eff}\ 1,2}^{(2)}$  
can be described in the following way. 
The projector $\Pi_0$ allows states of the form (\ref{ground}), so we start 
observing that the Hamiltonian vanishes on states of the following form
\begin{equation}\label{twosites}
\cW^{\uparrow}\site 1 \wb\cW^{\downarrow}~ \otimes~ \cW^{\uparrow}\site 2 
\wb\cW^{\downarrow} \qquad \text{or} \qquad 
\wb\cW^{\uparrow}\site 1 \cW^{\downarrow}~ \otimes~ \wb\cW^{\uparrow}\site 2 
\cW^{\downarrow}
\end{equation}
because the projectors in (\ref{secondord}) require orthogonal 
subspaces on different sites for the same type (e.g. up) of vectors. 
For example, this means that a state 
$v^{\uparrow}\site 1 \wb v^{\downarrow}~ \otimes~ v^{\uparrow}\site 2 
\wb v^{\downarrow}$ is killed by the two-sites Hamiltonian. 
We are left with states of the form 
\begin{equation}\label{twosites2}
\cW^{\uparrow}\site 1 \wb\cW^{\downarrow}~ \otimes~ \wb\cW^{\uparrow}\site 2 
\cW^{\downarrow} \qquad \text{or} \qquad 
\wb\cW^{\uparrow}\site 1 \cW^{\downarrow}~ \otimes~ \cW^{\uparrow}\site 2 
\wb\cW^{\downarrow}
\end{equation}
on which the parenthesis of projectors 
$\big(\pi^{\uparrow}_j\,\pi^{\downarrow}_{j+1}+\pi_j^{\downarrow}\,\pi^{\uparrow}_{j+1}
\big)$
acts as the identity. 
A state in one of the spaces (\ref{twosites2}) is respectively of the form 
\begin{equation}\label{twosites3}
v\site 1 \wb w ~ \otimes~ \wb v \site 2 w \quad , \quad
\wb v\site 1 w ~ \otimes~ v \site 2 \wb w
\end{equation}
on which we have respectively
\begin{eqnarray}
\big(1+P^{\uparrow}_{j,j+1} P^{\downarrow}_{j,j+1}\big)\  
v\site 1 \wb w ~ \otimes~ \wb v \site 2 w &=&
v\site 1 \wb w ~ \otimes~ \wb v \site 2 w+(-1)^{([v] +[\wb w])([\wb v]+[w])}\ 
\wb v\site 1  w ~ \otimes~  v \site 2 \wb w \nonumber \\
&& \label{twosites4} \\
\big(1+P^{\uparrow}_{j,j+1} P^{\downarrow}_{j,j+1}\big)\  
\wb v\site 1 w ~ \otimes~ v \site 2 \wb w &=&
\wb v\site 1 w ~ \otimes~ v \site 2 \wb w+(-1)^{([v] +[\wb w])([\wb v]+[w])}\ 
v\site 1 \wb w ~ \otimes~ \wb v \site 2 w \nonumber
\end{eqnarray}
In matricial form, the two-sites Hamiltonian has one of the two block-diagonal structures
\begin{equation}\label{twosites5}
B_{-}=\begin{pmatrix} 1& -1\\-1 &1 \end{pmatrix} \qquad \text{or}
\qquad B_{+}=\begin{pmatrix} 1& 1\\1 &1 \end{pmatrix} \,,
\end{equation}
all other entries being zero \cite{XX}. Both the blocks have eigenvalues 0 and 2.
The multiplicity depends on the actual model under examination.  
In the Hubbard model, the effective Hamiltonian acting on the singly occupied states
reduces to the block $B_{-}$ only
\begin{equation}
H^{(2)}_{\text{eff }j,j+1}=1-4\,\mathbf{S}_j\cdot \mathbf{S}_{j+1}=
2\begin{pmatrix} 0 &0& 0& 0\\0 & 1&-1&0\\
0&-1&1&0\\0&0&0&0\end{pmatrix}
\end{equation}
where $\mathbf{S}_j=(S^x,S^y,S^z)$ on site $j$ are the spin vectors of the Heisenberg 
model.

\subsection{Fourth order Hamiltonian}
The fourth order Hamiltonian 
\begin{gather}\label{fourthord}
H_{\text{eff}}^{(4)} = \sum_j H^{(4)}_{\text{eff}\ j,j+1,j+2}=\\
=\frac{1}{32} \sum _j \left[ 
\big(1+2P^{\uparrow}_{j,j+1} P^{\downarrow}_{j,j+1}+ 
P^{\uparrow}_{j+1,j+2} P^{\downarrow}_{j+1,j+2} P^{\uparrow}_{j,j+1} P^{\downarrow}_{j,j+1}\big) 
\big(\pi^{\downarrow}_j\pi^{\uparrow}_{j+1}\pi_{j+2}^{\uparrow}+
\pi^{\uparrow}_j\pi^{\downarrow}_{j+1}\pi_{j+2}^{\downarrow}\big)+ \right.\nonumber\\
\left.+\big(1+ 2P^{\uparrow}_{j+1,j+2} P^{\downarrow}_{j+1,j+2}+ 
P^{\uparrow}_{j,j+1} P^{\downarrow}_{j,j+1} P^{\uparrow}_{j+1,j+2} P^{\downarrow}_{j+1,j+2}\big) 
\big(\pi^{\downarrow}_j\pi^{\downarrow}_{j+1}\pi_{j+2}^{\uparrow}+
\pi^{\uparrow}_j\pi^{\uparrow}_{j+1}\pi_{j+2}^{\downarrow}\big)+\right.\nonumber\\
\left. +2 \big(2+P^{\uparrow}_{j,j+1} P^{\downarrow}_{j,j+1}+
P^{\uparrow}_{j+1,j+2} P^{\downarrow}_{j+1,j+2}\big) 
\big(\pi^{\uparrow}_j\pi^{\downarrow}_{j+1}\pi_{j+2}^{\uparrow}+
\pi^{\downarrow}_{j}\pi_{j+1}^{\uparrow}\pi^{\downarrow}_{j+2}\big) \right]  \Pi_0\nonumber
\end{gather}
is composed by a three-sites Hamiltonian density. It acts generically on states 
$\pi^{\varepsilon_1}_j\pi^{\varepsilon_2}_{j+1}\pi_{j+2}^{\varepsilon_3}$, 
where $\varepsilon=\pm1$ 
indicates respectively $\uparrow$ and $\downarrow$.
It cannot mix states with different values of $\varepsilon_1+\varepsilon_2+\varepsilon_3$; 
if that 
sum is $\pm 3$, its action is zero. The other two possible values are $\pm 1$,
on which it acts independently.

The second order Hamiltonian can be put in a three-sites density form by averaging 
on neighboring sites
\begin{equation}
\frac12 (H_{\text{eff}\ j,j+1}^{(2)}+H_{\text{eff}\ j+1,j+2}^{(2)})
\end{equation}
therefore we can evaluate the eigenvalues of a three-sites Hamiltonian
formed by
\begin{equation}
\mathcal{H}_{\text{eff}\ j,j+1,j+2}=
\frac12 \Big( H_{\text{eff}\ j,j+1}^{(2)}+H_{\text{eff}\ 
j+1,j+2}^{(2)}\Big)
+\frac{1}{U^2} H_{\text{eff}\ j,j+1,j+2}^{(4)}
\end{equation}
by action on states as in (\ref{twosites4}).
The possible eigenvalues are (up to corrections of order $U^{-4}$) 
the same in all cases
\begin{equation}
\mbox{Eigen}(\mathcal{H}_{\text{eff}\ 
j,j+1,j+2})=\left\{0,1,3\Big(1+\frac{1}{16\,U^2}\Big)\right\}
\end{equation}
with multiplicities that depend on the specific model under consideration.

\section{Conclusion}
We have defined in a very general way Hubbard models with arbitrary
symmetry. The basic ingredients are a vector space, which defines the
representation space on each site, and two projectors, which separate the
particles into two classes that behave in a different way. The scattering
matrix, as well as the energies, have been computed on very general ground.
The general form of Bethe ansatz equations remains to be computed, although
some of them are given here. Of course applications of these models to
condensed matter physics and/or AdS/CFT correspondence are of first
importance. In this regard, we have given in appendices a general procedure
to include a Aharonov-Bohm phase and some hints towards the definition of
integrable bosonic Hubbard models. This latter feature is especially
crucial, in particular for applications in string theory regarding the
AdS/CFT correspondence. Indeed, dealing with models with $psu(4|4)$
symmetry, one is lead to consider some subsectors of the theory containing
both fermionic and bosonic particles. The study of condensed matter models
with bosonic content on an integrable point of view also requires progress
in this direction. Finally, let us stress that the input of boundaries,
which play a great role in condensed matter models, may be a worthwhile
extension of this work, see e.g. \cite{shiwa} for the one-dimensional
Hubbard model with integrable boundaries.

\appendix
\section{Towards integrable bosonic Hubbard models\label{sec:XXbos}}
Our construction can in principle be applied to infinite dimensional 
vector spaces, leading to possible bosonic integrable Hubbard models. 
However, for such a purpose, one needs to construct a trace operator 
on the space $\cV$, a task that is not guaranteed when $\cV$ is 
infinite dimensional. To simplify the presentation, we work on XX 
models, but the procedure leading to Hubbard models can be applied in 
the same way we did for finite-dimensional vector spaces.
\subsection{$R$-matrices associated to Fock space}
To illustrate the problems encountered in the 
infite dimensional case, we focus on the case where $\cV$ is the Fock 
space $\cF$, based on oscillators $(b,b^\dag)$ with 
$[b\,,\,b^\dag]=1$. We denote by $|0>$ the Fock vacuum, and 
$|n>=(b^\dag)^n\,|0>$, $n=0,1,2,\ldots$, and by $\wh N=b^\dag\,b$ the 
number operator:
\beq
\wh N\,\vert n> = n\,\vert n>\,.
\eeq
 The dual vectors are noted $<n\vert$:
\beq
<m\vert\,.\,\vert n>\equiv <m\vert n>=\delta_{n,m}\,.
\eeq
We also introduce the subsets 
\beq
\cF_{N}=\Big\{|n>\,,\ n=0,1,2,\ldots,N\Big\}\subset\cF\,, \ 
N=0,1,2,\ldots
\eeq
The permutation operator is given by
\begin{equation}
P_{12}=\sum_{n,m=0}^\infty |n><m|\otimes |m><n|
\end{equation}
We present two examples of projectors. 

\paragraph{Even-odd projectors:}
One can choose as projectors 
\beq
\pi^{ev}=\frac{1}{2}(1+(-1)^{\wh N})\mb{and}
\wb\pi^{ev}=\II-\pi^{ev}\equiv\pi^{odd}=\frac{1}{2}(1-(-1)^{\wh N})\,,
\eeq
 which project on even and odd 
particle number eigenspaces $\cF_{ev}$ and $\cF_{odd}$. They obviously 
commute with $\wh N$. The parity operator is
$C=(-1)^{\wh N}$. Then, the corresponding XX $R$-matrix reads
\beq
R(\lambda) = \cos(\frac{\lambda}{2})\,\Big\{
\cos(\frac{\lambda}{2})\,P_{12} + 
\sin(\frac{\lambda}{2})\,\II\otimes\II\Big\}
-\sin(\frac{\lambda}{2})\,(-1)^{\wh N_{1}+\wh N_{2}}\Big\{
\sin(\frac{\lambda}{2})\,P_{12} + 
\cos(\frac{\lambda}{2})\,\II\otimes\II\Big\}
\eeq
where $\wh N_{1}=\wh N\otimes\II$ and $\wh N_{2}=\II\otimes \wh N$. 
\paragraph{Small modes projector:}
For any number $\ell\in\ZZ_{+}$, one can also take as projectors 
\beq
\pi^{\leq\ell}=\sum_{n=0}^\ell|n><n| \mb{and} 
\wb\pi^{\leq\ell}=\II-\pi^{\leq\ell}\equiv\pi^{>\ell}=\sum_{n>\ell}|n><n|\,.
\eeq
They also commute with $\wh N$. The parity operator reads
\beq
C=\sum_{n=0}^\ell|n><n|-\sum_{n=\ell+1}^\infty|n><n|\,.
\eeq
\paragraph{Properties of the $R$-matrix:}
The different types of projectors lead to different 
types of $R$-matrices. From the general treatment done in section 
\ref{sec:univR-XX}, one already knows that these $R$-matrices obey 
the theorem \ref{theo:univR-XX}, in particular the Yang-Baxter 
equation. 
Then, one directly constructs a monodromy matrix
\begin{equation}
\cL_{0<1\ldots L>}(\lambda) = R_{01}(\lambda)\,R_{02}(\lambda)\cdots
R_{0L}(\lambda)
\end{equation}
which obeys the relation
\begin{equation}
R_{00'}(\lambda-\mu)\, \cL_{0<1\ldots L>}(\lambda)\, 
\cL_{0'<1\ldots L>}(\mu) =
\cL_{0'<1\ldots L>}(\mu) \, \cL_{0<1\ldots L>}(\lambda)\,
R_{00'}(\lambda-\mu)\,.
\end{equation}
However, to get a transfer matrix leading to an integrable model, one 
first needs to define the trace operator: we discuss it in the next 
section.

We recall that the same construction is valid for the Hubbard $R$-matrices 
that one could build 
by coupling two XX models, as done in section \ref{sec:univR-Hub}. 

\subsection{Trace operator and transfer matrix}
One can define a trace operator for operators $\cO$ such that 
$[\cO\,,\,\wh N]=0$ in the following way. 
Since such operators preserve the subsets
$\cF_{N}$, which are finite dimensional spaces, their trace on 
$\cF_{N}$
\beq
tr_{N}\cO=\sum_{n=0}^N <n|\cO|n>
\eeq
 is well-defined and cyclic\footnote{In fact it is true for any 
operator $\cO$
obeying $\cO(\cF_{N})\subset\cF_{N}$.}. The trace on $\cF$ is then 
defined by the inductive limit 
\beq
tr\,\cO=\lim_{N\to\infty}\frac{1}{N}\,tr_{N}\,\cO\,. 
\eeq
It is cyclic and such that 
\beq
tr\II=1\mb{;} tr\,\pi^{ev}=tr\,\wb\pi^{ev}=\half
\mb{;} tr\,\pi^{\leq\ell}=0 \mb{and} tr\,\wb\pi^{\leq\ell}=1\,.
\eeq
In the same way, the permutation operator commutes with 
$\wh N_{1}+\wh N_{2}$, so that $tr_{12}P_{12}$ is 
well-defined and cyclic. However, to define the transfer matrix, 
we need to use the partial trace 
$tr_{1}P_{12}$ which is ill-defined because $P_{12}$ does not commute 
with $\wh N_{1}$. For instance, it is easy to see
\beq
tr_{N_{1}}(P_{12}\,P_{13})\neq tr_{N_{1}}(P_{13}\,P_{12})
\label{eq:noncycl}
\eeq
where $tr_{N_{1}}$ is the operator $tr_{N}$ in the space 1.
Equation (\ref{eq:noncycl}) just shows that the partial trace 
is not cyclic. As a consequence, one cannot prove that transfer 
matrices with different spectral parameters commute, and the 
integrability of the model is not guaranteed.

\section{Twisted version of XX and Hubbard models\label{sec:twist}}
The Hubbard models we have constructed depend on a single free 
parameter $U$. We present here a construction that allows us to introduce 
more parameters. In particular, we will obtain an Hermitian 
Hamiltonian that depends on phases that can be identified with a 
Aharonov-Bohm phase. Again, we present in detail the construction 
for XX models, and just sketch the generalization to Hubbard models.

\null

We start with an universal XX model defined on the vector space $\cV$, 
with a projector $\pi$ such that $\pi(\cV)=\cW$. 
\begin{definition}
To any projection $\pi$, a refinement is a decomposition
$$
\pi=\oplus_{j=1}^d \pi^{(j)} \mb{with} 
\pi^{(j)}\pi^{(j')}=\delta_{j,j'}\,\pi^{(j)}\qquad \forall\,j,j'=1,\ldots,d
$$
\end{definition}
For each refinement of $\pi$ and $\wb\pi$, we introduce
$$
\cF_{12} = \II\otimes\II +\sum_{a=1}^d\sum_{\bara=1}^{\bar d} 
(q_{a\bara}-1)\,\pi^{(a)}\otimes \pi^{(\bara)} 
$$
where $q_{a\bara}$ are some non-zero complex numbers. Note that 
$q_{a\bara}\neq 0$ ensures that $\cF_{12}$ is invertible. Its inverse 
reads
$$
\cF_{12}^{-1} = \II\otimes\II -\sum_{a=1}^d\sum_{\bara=1}^{\bar d}
\frac{q_{a\bara}-1}{q_{a\bara}}\,\pi^{(a)}\otimes \pi^{(\bara)}  
$$

The twisted version of the XX model is defined by the R-matrix
$$
\wh{R}_{12}(\lambda) = \cF_{12}\,R_{12}(\lambda)\,\cF_{21}^{-1}
$$
where $R_{12}(\lambda)$ is the matrix (\ref{def:univRXX}). We remind that 
it depends on the choice of the projector $\pi$.
\begin{property}\label{prop:Rhat}
The R-matrix $\wh{R}_{12}(\lambda)$ can be rewritten as 
\begin{eqnarray}
\wh{R}_{12}(\lambda)  &=& 
\wh{\Sigma}_{12}\,\sin\lambda +
\Big(\Sigma_{12} + (\II\otimes\II-\Sigma_{12})\,\cos\lambda
\Big)\,P_{12} \label{def:RXXhat}\\
\wh{\Sigma}_{12} &=& \cF_{12}\,\Sigma_{12}\,\cF_{21}^{-1}
=\sum_{a=1}^d\sum_{\bara=1}^{\bar d} \left(
q_{a\bara}\,\pi^{(a)}\otimes \pi^{(\bara)} 
+\frac{1}{q_{a\bara}}\,\pi^{(\bara)}\otimes \pi^{(a)}  \right)
\label{def:Sigmahat}
\end{eqnarray}
It obeys all the properties stated in theorem \ref{theo:univR-XX}, but the 
symmetry one.
\end{property}
\prf
The expressions (\ref{def:RXXhat}) and (\ref{def:Sigmahat}) follow 
from direct calculations.\\
Remarking that for any diagonal matrix $\cD_{12}$, we have 
the property
$$
\cD_{13}\,\cD_{23}\,R_{12}(\lambda) = 
R_{12}(\lambda)\,\cD_{13}\,\cD_{23}
$$
the YBE for $\wh{R}_{12}(\lambda)$ is deduced from the YBE for 
${R}_{12}(\lambda)$. For instance, starting from the l.h.s. and using 
the notation $\lambda_{ij}=\lambda_{i}-\lambda_{j}$:
\begin{eqnarray*}
&&\wh{R}_{12}(\lambda_{12})\,\wh{R}_{13}(\lambda_{13})
\,\wh{R}_{23}(\lambda_{23})\ =\  
\cF_{12}\,R_{12}(\lambda_{12})\,\cF_{21}^{-1}\,
\cF_{13}\,R_{13}(\lambda_{13})\,\cF_{31}^{-1}\,
\cF_{23}\,R_{23}(\lambda_{23})\,\cF_{32}^{-1}\\
&=&\cF_{12}\,R_{12}(\lambda_{12})\,\cF_{13}\,
\cF_{23}\,\cF_{23}^{-1}\,\cF_{21}^{-1}\,
R_{13}(\lambda_{13})\,\cF_{31}^{-1}\,
\cF_{23}\,R_{23}(\lambda_{23})\,\cF_{32}^{-1}\\
&=&\cF_{12}\,\cF_{13}\,\cF_{23}\,R_{12}(\lambda_{12})\,
R_{13}(\lambda_{13})\,\cF_{31}^{-1}\,
\cF_{23}^{-1}\,\cF_{21}^{-1}\,\cF_{23}\,
R_{23}(\lambda_{23})\,\cF_{32}^{-1}\\
&=& \cF_{12}\,\cF_{13}\,\cF_{23}\,R_{12}(\lambda_{12})\,
R_{13}(\lambda_{13})\,R_{23}(\lambda_{23})\,\cF_{31}^{-1}\,
\cF_{21}^{-1}\,\cF_{32}^{-1}
\end{eqnarray*}
The r.h.s. is treated in the same way.\\
Regularity is obtained as follows
$$
\wh{R}_{12}(0) = \cF_{12}\,R_{12}(0)\,\cF_{21}^{-1} = 
\cF_{12}\,P_{12}\,\cF_{21}^{-1} =\cF_{12}\,\cF_{12}^{-1}\,P_{12} =
P_{12} 
$$
Similar calculations lead to the other properties.
\finprf
Property \ref{prop:Rhat} ensures that models based on twisted 
$R$-matrices are also integrable and possess a local Hamiltonian.
Indeed, the corresponding Hamiltonian reads
\begin{equation}
\wh{H}(q_{a\bara})=\sum_{j=1}^{L} \wh{H}_{j,j+1}(q_{a\bara})
\mb{with} \wh{H}_{j,j+1}=
P_{j,j+1}\,\wh{\Sigma}_{j,j+1}(q_{a\bara})
\label{eq:XXHamhat}
\end{equation}
Its hermiticity depends on the parameters $q_{a\bara}$.
From the calculation
\ben
\wh{H}(q_{a\bara}) &=& \sum_{j=1}^{L}
P_{j,j+1} \wh{\Sigma}_{j,j+1}(q_{a\bara})\\
\wh{H}(q_{a\bara})^\dag &=& \sum_{j=1}^{L}
\wh{\Sigma}_{j,j+1}(q_{a\bara}^{*})\,P_{j,j+1}= \sum_{j=1}^{L}
P_{j,j+1}\,\wh{\Sigma}_{j+1,j}(q_{a\bara}^{*})
\een
and the identity
$$
\wh{\Sigma}_{j+1,j}(q_{a\bara})= 
\wh{\Sigma}_{j,j+1}(\frac{1}{q_{a\bara}})
$$
one deduces that $\wh{H}$ is hermitian when the parameters are 
phases: 
$$
q_{a\bara}=e^{i\theta_{a\bara}}\,,\quad\theta_{a\bara}\in\RR\,.
$$

Chosing the parameters $q_{a\bar a}$ to be phases, we have in this way a general Hermitian 
Hamiltonian with Aharonov-Bohm phases on each site of the model.

\paragraph{Twisted Hubbard models:}
The same construction can be done for Hubbard models, with now
$$
\wh{R}^{\uparrow\downarrow}_{12}(\lambda) = 
\cF^{\uparrow\downarrow}_{12}(\bq^\uparrow,\bq^\downarrow)\,R^{\uparrow\downarrow}_{12}(\lambda)
\,\left(\cF^{\uparrow\downarrow}_{21}(\bq^\uparrow,\bq^\downarrow)\right)^{-1}
\mb{with} \cF^{\uparrow\downarrow}_{12}(\bq^\uparrow,\bq^\downarrow)=
\cF^{\uparrow}_{12}(\bq^\uparrow)\,\cF^{\downarrow}_{12}(\bq^\downarrow)\,.
$$
Since $\cF$ and $C$ are diagonal, the new $R$-matrix reads
\begin{equation}
\wh R^{\uparrow\downarrow}_{12}(\lambda_{1},\lambda_{2}) =
\wh R^{\uparrow}_{12}(\lambda_{12})\,\wh R^{\downarrow}_{12}(\lambda_{12}) +
\frac{\sin(\lambda_{12})}{\sin(\lambda'_{12})} \,\tanh(h'_{12})\,
\wh R^{\uparrow}_{12}(\lambda'_{12})\,C^{\uparrow}_{1}\,
\wh R^{\downarrow}_{12}(\lambda'_{12})\,C^{\downarrow}_{1}
\end{equation}
where $\wh R^\eps(\lambda)$, $\eps=\uparrow,\downarrow$, have the 
form (\ref{def:RXXhat}). It leads to an Hamiltonian
\beq
\wh H= \sum_{j=1}^{L}\Big(
P^{\uparrow}_{j,j+1} \wh{\Sigma}^{\uparrow}_{j,j+1}(q^{\uparrow}_{a\bara})
+P^{\downarrow}_{j,j+1} \wh{\Sigma}^{\downarrow}_{j,j+1}(q^{\downarrow}_{a\bara})
+U\,C^{\uparrow}_{j}\,C^{\downarrow}_{j}\Big) 
\eeq
that is hermitian as soon as the parameters $q_{a\bar a}^\eps$ are phases.

\end{document}